\documentclass[twocolumn,tighten]{aastex63}

\begin{document}

\title{Searching for Molecular Outflows with Support Vector Machines: Dark Cloud Complex in Cygnus}

\author{Shaobo Zhang}
\author{Ji Yang}
\author{Ye Xu}
\author{Xuepeng Chen}
\author{Yang Su}
\author{Yan Sun}
\author{Xin Zhou}
\author{Yingjie Li}
\author{Dengrong Lu}
\affiliation{Purple Mountain Observatory, \& Key Laboratory for Radio Astronomy, Chinese Academy of Sciences, Nanjing  210023, China}

\correspondingauthor{Shaobo Zhang}
\email{shbzhang@pmo.ac.cn}

\begin{abstract}

We present a survey of molecular outflows across the dark cloud complex in the Cygnus region, based on 46.75~deg$^2$ field of CO isotopologues data from Milky Way Imaging Scroll Painting (MWISP) survey. A supervised machine learning algorithm, Support Vector Machine (SVM), is introduced to accelerate our visual assessment of outflow features in the data cube of $^{12}$CO and $^{13}$CO J = 1$-$0 emission. A total of 130 outflow candidates are identified, of which 77 show bipolar structures and 118 are new detections. Spatially, these outflows are located inside dense molecular clouds and some of them are found in clusters or in elongated linear structures tracing the underlying gas filament morphology. Along the line of sight, 97, 31, and 2 candidates reside in the Local, Perseus, and Outer arm, respectively. Young stellar objects as outflow drivers are found near most outflows, while 36 candidates show no associated source. The clusters of outflows that we detect are inhomogeneous in their properties; nevertheless, we show that the outflows cannot inject turbulent energy on cloud scales. Instead, at best, they are restricted to affecting the so called "clump" and "core" scales, and this only on short ($\sim$0.3~Myr) estimated timescales. Combined with outflow samples in the literature, our work shows a tight outflow mass-size correlation.

\end{abstract}


\section{Introduction} \label{sec:intro}

Molecular outflows are ubiquitous manifestations observed in star forming regions, as it is thought to be an indispensable process that every protostar undergoes \citep{bal83, shu87, fra14, bal16}. Since the discovery of the first bipolar structure by \citet{sne80}, numerous observational and theoretical studies have been carried out to reveal outflow as a key link in star forming process. As outflows not only record the histories of mass ejection from forming young stars, but can also be related to mass accretion to the protostars \citep{bal16}, as well as feedback to the environment \citep{elm04, sca04, mat07}.

Observations usually diagnose molecular outflows by high-velocity gas shown on molecular line wings, which show mono-polar, bipolar, or multi-polar localized structures \citep{wu04}. These molecular outflows present dimensions of up to several parsecs, wide mass ranges from $10^{-3}$ to $10^3$~M$_\odot$, and a typical kinetic energy of $10^{45}$~erg \citep{bal83, wu04}. On the other hand, theoretical studies proposed a wide range of outflow models concerning their launching mechanism \citep{shu00, pud07}, the formation of molecular outflows (summarized by \citealt{arc07}), and their feedback in self-regulating star formation \citep{nak07, fed14}. However, the models have undergone long-term debates, since no single model could explain all the features observed, especially for the outflows formed from clusters and massive stars, toward which there are a limited number of observations. Large-scale unbiased survey of outflows could move us forward in solving the arguments.

The traditional methods of outflow searching are usually based on reference sources of known star formation activity obtained from other observations \citep{bal83, dob01, hat07, li15}. Either infrared sources or dust cores were used as indicators for molecular outflows, which implants biases during source selection, and may limit the outflow study on cloud-scale. Unbiased Large-scale searching for molecular outflow became feasible in the past decade on account of mass data acquired from a number of molecular line surveys with (sub-)arcmin angular resolutions. CO surveys toward the Perseus, Taurus, and other regions were carried out using Five College Radio Astronomy Observatory (FCRAO) telescope, and James Clerk Maxwell Telescope (JCMT). Molecular outflows were manually searched in these regions, and hundreds of new outflows were detected \citep{hat09, arc10, cur10, nar12, li15, li18}. However, studies of manual identification are time consuming, non-repeatable, and involving subjective factors such as people's perception of a image.

There are a limited number of attempts in introducing computer-assisted methods to outflow searching. \citet{arc10} proposed a semi-automated procedure to visualize outflow candidates as spikes on three-dimensional isosurface. Based on such idea, \citet{li18} combined the isosurface with optically-thin lines and adopted clump-finding algorithm to search for the spikes. However, these approaches are still based on empirical criteria, as traditional methods, in which numbers of fine-turned thresholds are necessary to find matched candidates. In recent years, machine learning algorithms have been widely concerned in the field of pattern recognition, with its simplicity and preciseness. Relying on a set of pre-labeled samples rather than fixed criteria, they would be ideal tools to identify structures like outflow which are difficult to define explicitly.

In this paper, we present a survey of molecular outflows towards a complex of dark clouds in Cygnus based on a machine learning algorithm. A pixel-by-pixel searching procedure is established to identify gas emissions of outflows in CO data cubes. The structure of this paper is as follows. In Section~\ref{sec:data}, we describe the observations we made and distributions of molecular clouds in the data. The searching method for outflow features is given in Section~\ref{sec:method}. In Section~\ref{sec:result}, we present the outflow catalog and analyze physical properties of the detected samples. These properties comparing with those in the literatures, and feedback to the parent clouds are discussed in Section~\ref{sec:discuss}. The summary is then made in Section~\ref{sec:summary}.

\section{The Data} \label{sec:data}

\subsection{Observations}

As a part of the Milky Way Imaging Scroll Painting (MWISP) project\footnote{ \url{http://www.radioast.nsdc.cn/mwisp.php}} \citep{su19}, we observed a series of Lynds dark clouds in the Cygnus region connecting from L935 to IC~5146 with the Purple Mountain Observatory Delingha (PMODLH) 13.7~m telescope. $^{12}$CO~(1$-$0), $^{13}$CO~(1$-$0), and C$^{18}$O~(1$-$0) lines were observed simultaneously with the 3$\times$3-beam superconducting array receiver (SSAR) working in sideband separation mode, with a set of fast Fourier transform spectrometer (FFTS) employed \citep{sha12}. The observed region was separated into 187 cells of dimension 30\arcmin$\times$30\arcmin, and covered an total area of 46.75~deg$^2$. Each cell was mapped using the on-the-fly (OTF) observation mode, with at least two scans along the Galactic longitude and latitude directions to reduce the scanning effects. A total of ~600 hours were used for the observations from 2012 to 2016. The half-power beam width (HPBW) is 52\arcsec\ at 115~GHz, and the pointing accuracy is better than 5\arcsec.

The standard chopper wheel method \citep{uli76} was used to calibrate the antenna temperature. We divided the antenna temperature ($T_{\rm A}$) by the main beam efficiency ($B_{\rm eff}$) to obtain the main beam temperature ($T_{\rm mb}$). Different efficiencies were adopted, as $B_{\rm eff}$ varied within 44-52\% at 115~GHz and 48-56\% at 110~GHz during the five years of observations according to the annually status report of PMODLH. The calibration errors are estimated to be within 10\%. The OTF data were re-gridded to 30\arcsec\ pixels and then mosaicked to a FITS cube using the GILDAS software package \citep{pet05, gil13}. The L935 region has been mapped once in 2011 as a pilot survey (see \citealt{zha14} for details). We reprocessed these additional data following the same procedure, and included them while mosaicking. The resulting rms noise is 0.44~K for $^{12}$CO at the resolution of 0.16~km~s$^{-1}$, 0.25~K for $^{13}$CO and C$^{18}$O at 0.17~km~s$^{-1}$. The L935 region presents a lower noise level of 0.23~K and 0.15~K, respectively.


\subsection{Distributions of Molecular Clouds}

\begin{figure*}[htbp]
\centering
\includegraphics[trim=0 40 0 0, width=0.75\textwidth]{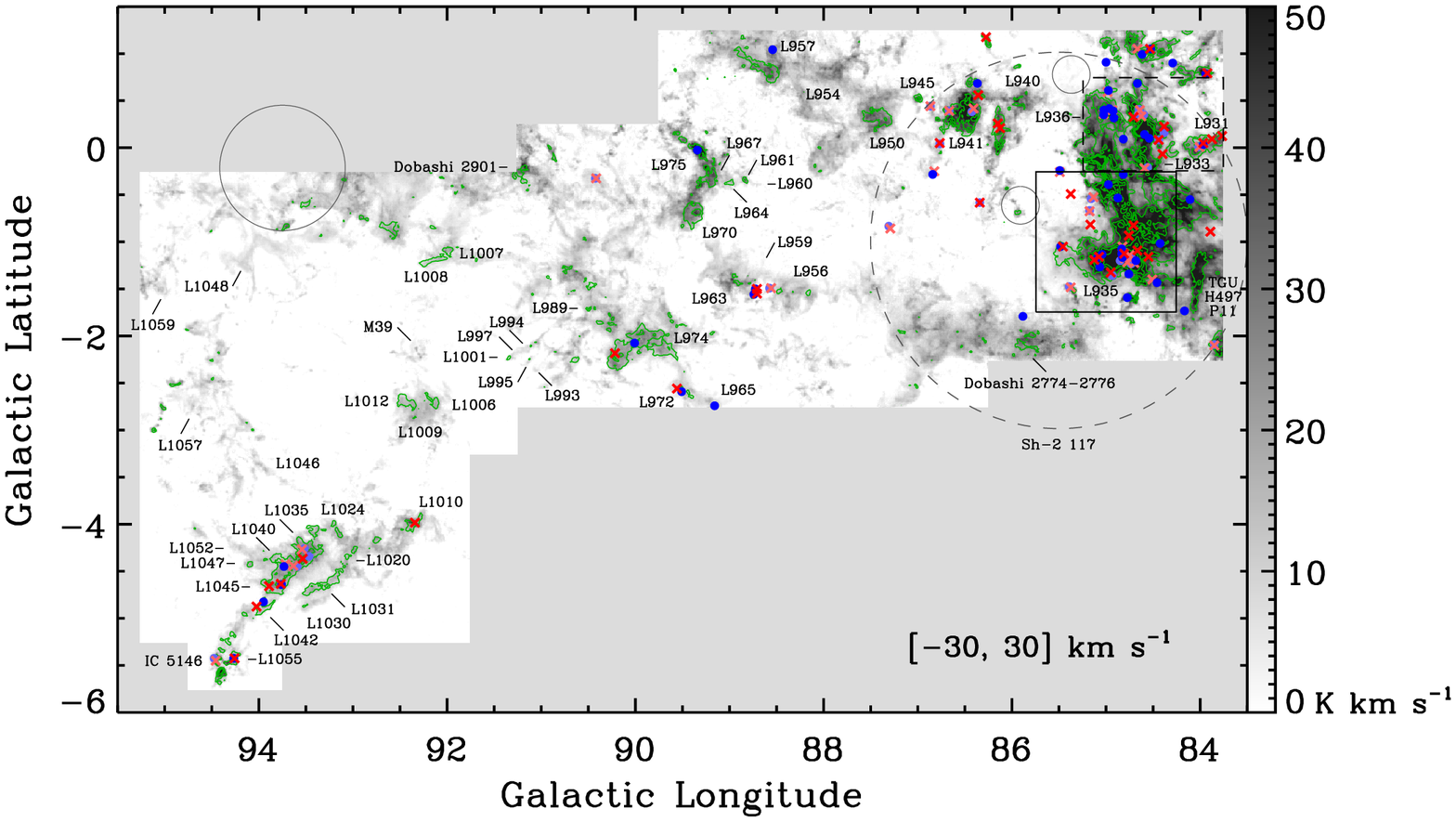} \\
\includegraphics[trim=0 40 0 0, width=0.75\textwidth]{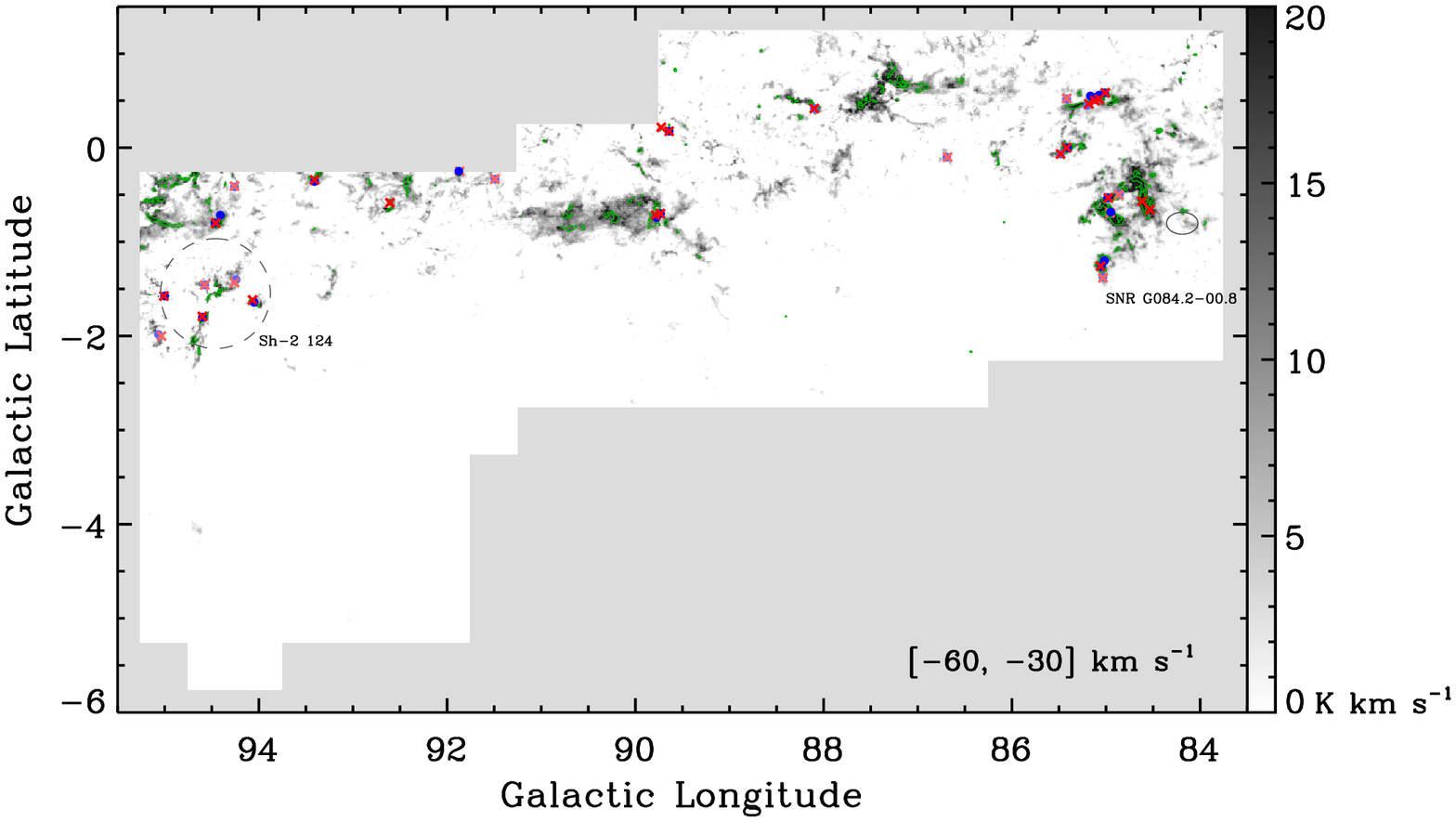} \\
\includegraphics[trim=0 0 0 0, width=0.75\textwidth]{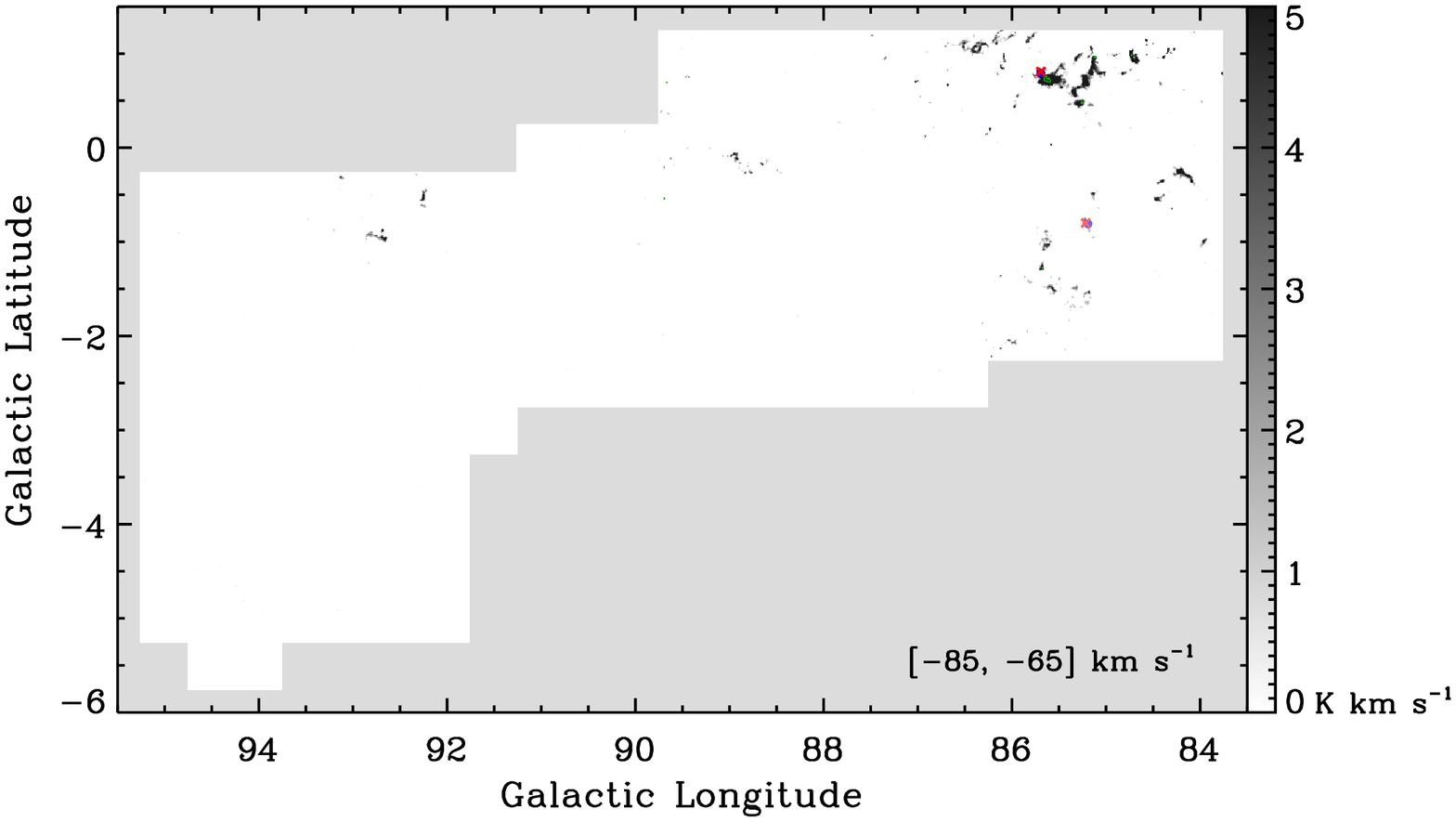}
\caption{Integrated intensity map over three velocity ranges: $-30 \sim 30$~km~s$^{-1}$ (top), $-60 \sim -30$~km~s$^{-1}$ (middle), and $-85 \sim -65$~km~s$^{-1}$ (bottom). The background gray-scale image is the integrated intensity map of $^{12}$CO, while the green contours represent the integrated intensity map of $^{13}$CO from $3\sigma$ at intervals of $5\sigma$. Dark nebulae associated molecular clouds are labeled on the map \citep{lyn62, dob05, dob11}, Circles indicate H~{\scriptsize II} regions (dashed circles) and supernova remnants (solid circles), whose positions and sizes are cataloged in \citet{sha59, kot01, gre14}. Outflows identified in this work are marked with blue dots and red crosses, which represent the blueshifted and redshifted lobes, respectively. The dark and light symbols indicate the outflow of grade A and B, respectively. Two boxes in the upper panel show the training region, while the dashed box outlines the validation set for cross-validation. \label{fig:guidemap}}
\end{figure*}

The spectra in our mapping region present three main velocity components, which correspond to the Local arm ($-30 \sim 30$~km~s$^{-1}$), the Perseus arm ($-60 \sim -30$~km~s$^{-1}$), and the Outer arm ($-85 \sim -65$~km~s$^{-1}$). The distributions of $^{12}$CO and $^{13}$CO in three components are shown in Figure~\ref{fig:guidemap}. $^{12}$CO emission is bright, and presents filamentary or extended structures of different spatial scales in three components throughout the mosaic, while $^{13}$CO presents condensations that follow the distribution of $^{12}$CO. C$^{18}$O is not presented in the map as its emission only appears around a few peak positions of $^{13}$CO.

Most of the molecular clouds in the Local arm are associated with dark nebulae \citep{lyn62, dob05, dob11}. The brightest portions among the clouds are the L933, L935, L936, L941, L975, and L1055 regions. A large H~{\scriptsize II} region, Sh-2 117, locates in the west of our mapping region, where associated cavity structures are found surrounding the ionized gas \citep{bal80, zha14}. Several Planck cold clumps \citep{pla16} locate within the filamentary clouds extending from L935 to IC~5146, which suggests a relatively quiescent environment. We note there is a large shell structure appearing to the north of IC~5146 centered at L1012 with a diameter of $\sim$4\arcdeg, which closely matches a far-infrared loop identified by \citet{kis04}. Cometary and finger-like molecular clouds could be found on the loop near L1048, L1008, and L1001. The integrated intensity ratio of $^{12}$CO to $^{13}$CO is 6.8 in regions with $^{13}$CO intensity over 2.5~K~km~s$^{-1}$. Such ratio is similar to the result in \citet{dob94}.

Different from the distributions in the Local arm, clouds in the Perseus arm mainly locate at $b>2$\arcdeg. Clouds seem to be less extended in the Perseus arm than those in the Local arm. Most clouds with strong $^{12}$CO emission are actively forming stars since there are massive young stellar objects(YSOs) and H~{\scriptsize II} regions embedding as reported by \citet{urq09}. Two linked ring-like structures with consistent velocities can be spotted in the westernmost of the mapping region (hereafter G084.9$-$0.4). One is centered at $l=84.8\arcdeg, b=-0.6\arcdeg$ with a radius of $\sim12\arcmin$, and the other larger one to the north is peaked at a filamentary arc near $l=85.1\arcdeg, b=0.5\arcdeg$. Molecular clouds in the Outer arm seem to be more compact since diffuse molecular gas far away from us ($\sim$9~kpc) is difficult to detect due to beam dilution. Still, filamentary and shell structures could be revealed in the map.

\section{Searching Procedures} \label{sec:method}

Our goal is to locate localized high-velocity wings (LHWs) in the data cube as outflow candidates. Considering the large amount of unbiased survey data, we require an automatic detector to blindly search outflow features based on a set of manually identified samples. In this way, our task becomes a supervised learning problem, and involves the following steps.

\begin{enumerate}
\item Manually identify LHWs in a representative subset of data as training samples. The LHWs checked by eye should meet the criteria of outflows (Section~\ref{sec:train}).
\item Extract numerical feature vectors from all data. These features should be able to summarize the differences between the LHW and other emission, from the aspect of line profile and spacial morphology (Section~\ref{sec:feature}).
\item Train and optimize the learning model with training data using cross-validation. The model trained using a portion of the training samples should perform well in picking out LHWs from the rest samples (Section~\ref{sec:optimize}).
\item Apply the model to the rest test data set to obtain LHW candidates. The candidates should then be graded based on criteria in step 1, and paired into bipolar outflows if possible (Section~\ref{sec:predict}).
\end{enumerate}

\subsection{Support Vector Machines}

Among the several statistical methods, we adopted the Support Vector Machine (SVM, \citealt{vap95}) algorithm for its excellent performance in multi-dimensional data. The SVM algorithm is a supervised learning algorithm used for classification and regression analysis. The main objective of SVM is to maximize the margin between different classes. This method has been widely used and proved to be efficient in different kinds of topics in astronomy, such as galaxies classification \citep{hue11}, shocked gas search in supernova remnants \citep{bea11}, YSO classification \citep{mar16}, etc. In our study, we adopted the SVM with Gaussian radial basis function kernel as a classifier, which calculated the decision boundaries between manually selected LHWs and non-LHWs, and applied the boundaries to the samples of unknown classes. SVM$^{light}$ \citep{joa99} provides implementations of SVM algorithm in a wide range of interfaces\footnote{A package in C was adopted in this work. Other versions as Pysvmlight and PySVMLight in Python; CRAN package klaR in the R statistical software environment; mex-svm in Matlab; and other interfaces are available at \url{http://http://svmlight.joachims.org}}. We wrote a set of procedures to invoke the package in IDL, and to visualize the candidates with GUIs.


\subsection{Training Set} \label{sec:train}

\begin{figure}[htbp]
\centering
\includegraphics[width=0.85\linewidth]{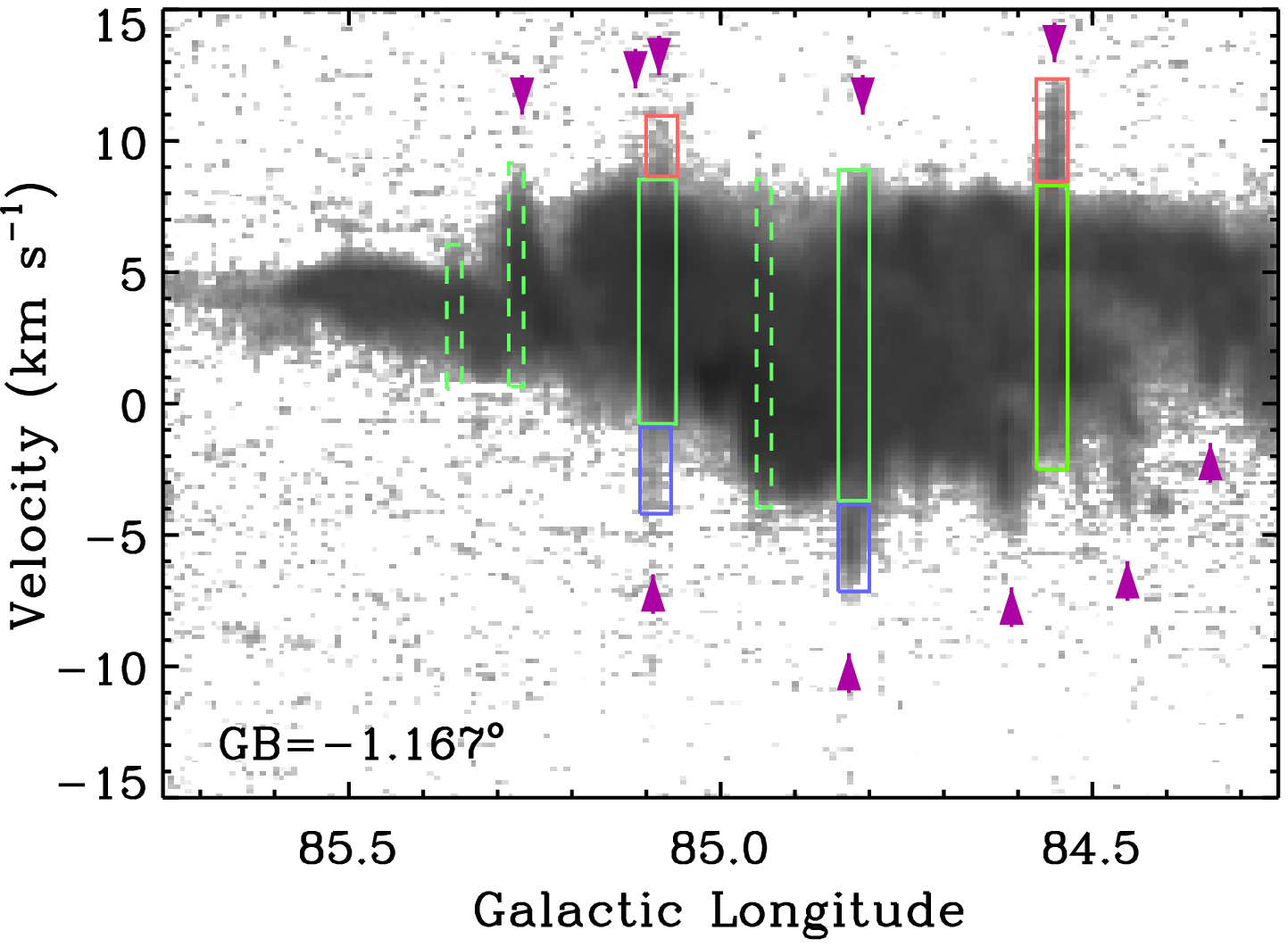}
\caption{$l$-$v$ slice through the $^{12}$CO data cube at $b=-1.167$\arcdeg. The figure demonstrates how the training data set was manually labeled. Purple triangles mark the velocity bumps for further check. Blue and red boxes mark the blue and red lobes of selected LHWs. Green boxes are the pixels labeled as non-LHW class, among which the dashed ones are the randomly selected spectra. \label{fig:traindemo}}
\end{figure}

For a supervised learning task, a training set is needed for the algorithm to infer the boundaries between different classes in feature space. The accuracy of classification can be improved by choosing a well labeled training set that represents the properties of the classes. We extracted two $^{12}$CO cubes in L935/936 region with velocity ranging from -15 to 15~km~s$^{-1}$ as the training data (boxes in Figure~\ref{fig:guidemap}). The training data account for $\sim$2\% of all the data set. In order to derive a representative and complete set that contains diverse LHWs, we run a blind search for velocity bumps in the Galactic longitude-velocity ($l$-$v$) slice of each Galactic latitude (as demonstrated in Figure~\ref{fig:traindemo}). Each bump was then checked in the $b$-$v$ map of each Galactic longitude. Their positions with the longest velocity extension were recorded as the top of the bumps. The search provided 315 bumps as the preliminary candidates for LHWs.

However, most of the candidates may not be LHWs, because velocity gradients, small clumps, and filamentary structures may all lead to similar bump features. To obtain a reliable LHW sample set, we checked each bump by eye with three diagrams and kept those that matched the corresponding criteria:

\begin{itemize}
\item We plotted the $^{12}$CO and $^{13}$CO spectra averaged within a 1\arcmin$\times$1\arcmin\ box centered at the top of each bump. LHWs should have line wing emission in excess of a Gaussian profile on the $^{12}$CO spectra but no $^{13}$CO component over $3 \sigma$ in the excess part.
\item We plotted the $^{12}$CO integrated intensity maps within a velocity range of 2~km~s$^{-1}$, which covers each bump with one end at the top of it. The emission over $3\sigma$ should be compact within a map size of 10\arcmin.
\item We plotted four position-velocity diagrams slicing through the tip of each bump at position angles of 0\arcdeg, 45\arcdeg, 90\arcdeg, and 135\arcdeg, respectively. The LHWs are expected to present bump on contours with velocity shift over 2~km~s$^{-1}$ in all four p-v maps.
\end{itemize}

Applying these criteria resulted in 35 LHWs, 18 of which are blueshifted lobes. Since we intended to do a pixel-by-pixel classification of the data, the LHWs were then disassembled into pixels. We extracted 7702 pixels in total in the line wings as LHW class of training set. Two types of pixels on the other hand were extracted as the non-LHW class of training set: (1) pixels on the spectra of LHWs but away from line wings. (2) pixels on the spectra at random selected positions away from LHWs. Finally, 60609 pixels were extracted for the non-LHW class.

\subsection{Feature Extraction} \label{sec:feature}

For a successful learning process, it is vital to choose a set of appropriate features which could tell the difference between the pixels of two classes. An advantage of SVM is that the algorithm performs stable even in a high dimensional feature space. For our data, the $^{12}$CO emissions are usually optically thick, thus we measured the velocity and line width of each component from $^{13}$CO lines. Five features were then extracted at each pixel with intensity over $3\sigma$ in $^{12}$CO data according to the following methods.

\begin{enumerate}
\item $T_{inner} / T_{peak}$, where $T_{inner}$ is the intensity averaged within a $3\times 3\times 7$ pixel sub-cube in the position-position-velocity space, and $T_{peak}$ is the peak intensity of $^{12}$CO spectra at the same spatial position. The sub-cube is centered at the pixel waiting to be classified, and corresponds to a $2.5\arcmin \times 2.5\arcmin \times 1~km~s^{-1}$ region, which guarantees that the LHWs larger than this size are not missing. This feature is expected to be lower in the line wings than near the line peak.

\item $| v_{\rm pixel} - v_{0} | / \Delta~v$, where $v_{\rm pixel}$ is the velocity of the pixel to be classified, $v_{0}$ and $\Delta v$ are the systemic velocity and line width estimated from $^{13}$CO component which the pixel belongs to. This feature represents the relative position of the pixel on the line profile, and is expected to be high in the line wings.

\item $s \times \Delta~v / T_{peak}$, where $s$ is the slope of a linear fitting to the averaged spectrum in the sub-cube, and $\Delta~v$ and $T_{peak}$ are added to scale the slope. We use this feature to prevent the emission between two adjacent components from being classified as LHW class. This feature is expected to be positive in most cases, but negative if another component appears.

\item $\Gamma_{\sigma_1,\sigma_2}$, the difference of Gaussians (DoG) of image, involves the image convolved with the subtraction of two Gaussian kernels of standard deviation $\sigma_1$ and $\sigma_2$ \citep{gon06}. It is a commonly used algorithm in image processing for blob detection. We choose this feature to describe the localization of LHWs, as the integration intensities of line wings are higher than those in the surrounding regions. The integration intensity map is derived within the same 7 channels as the sub-cube. We adopt $\sigma_1=2$ and $\sigma_2=4$ which means the feature is more sensitive in detecting structures with spatial dimension of $\sim$2-4\arcmin.

\item $T_{outer}/T_{peak}$, where $T_{outer}$ is the intensity averaged in the surrounding region with a radius of 4\arcmin, where the sub-cube is masked in the calculations. The feature further constrains the emission around the LHWs, and can help us to locate the spatial peak position of a LHW, where this feature is expected to be lower than the first feature.
\end{enumerate}

\begin{figure}[t]
\centering
\includegraphics[width=0.3\linewidth]{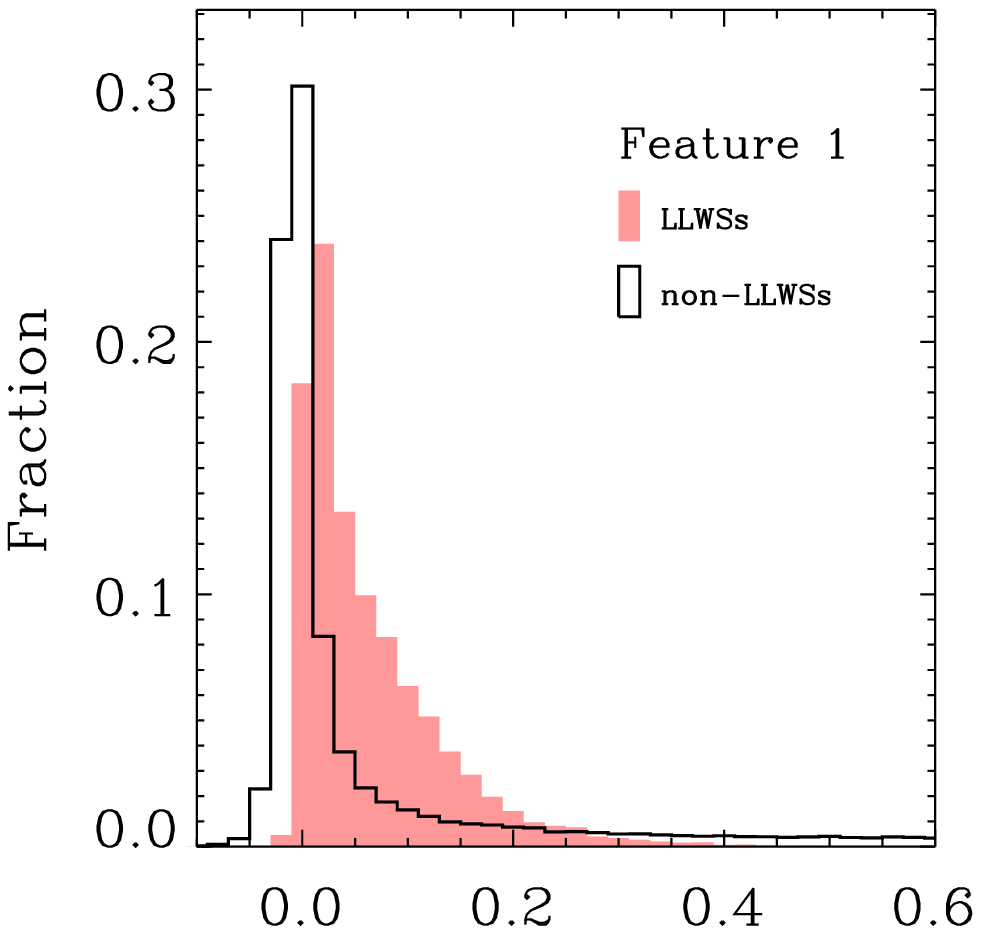}
\includegraphics[width=0.3\linewidth]{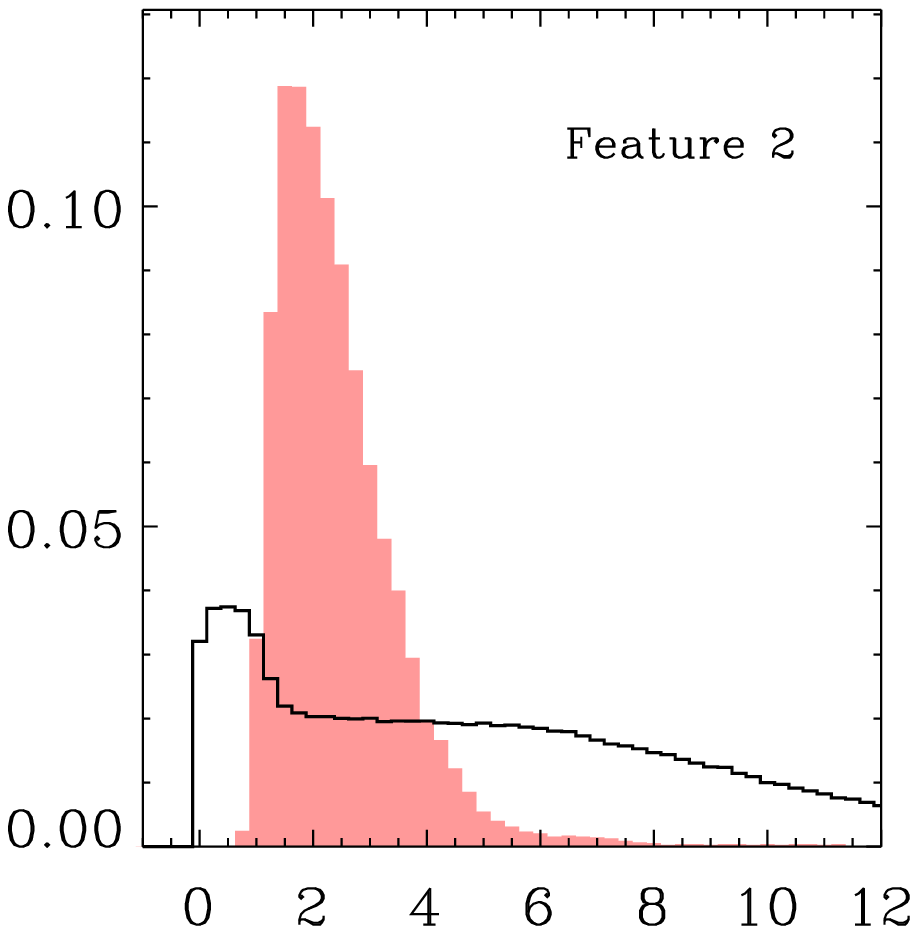}
\includegraphics[width=0.3\linewidth]{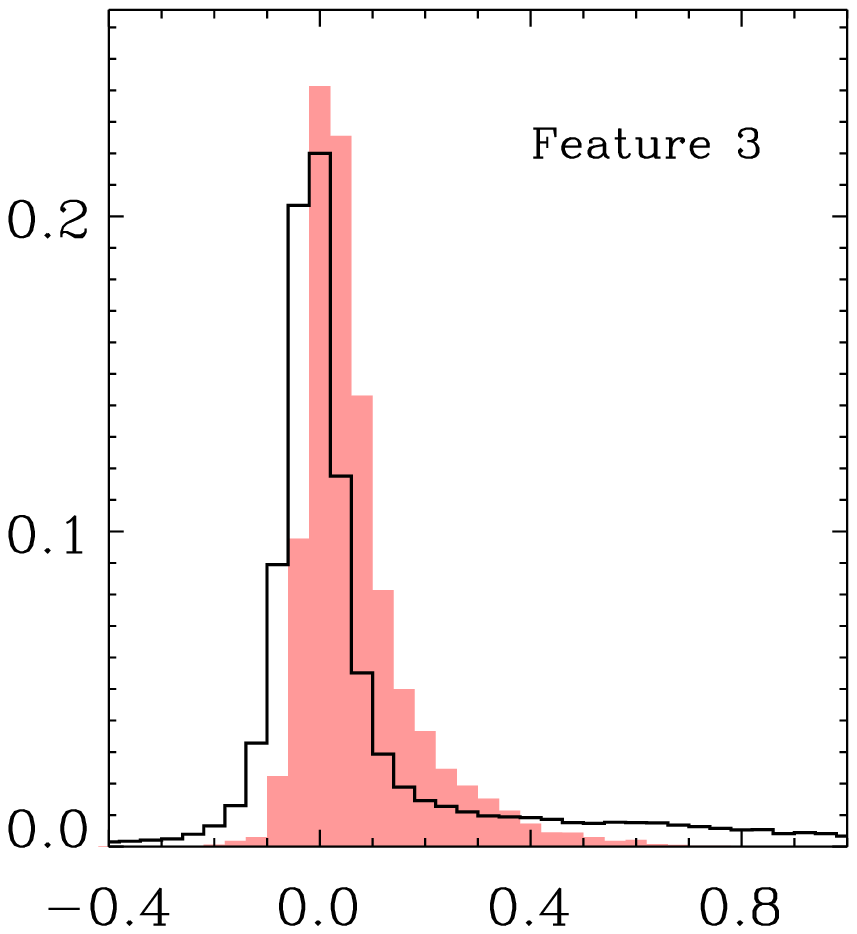} \\
\includegraphics[width=0.3\linewidth]{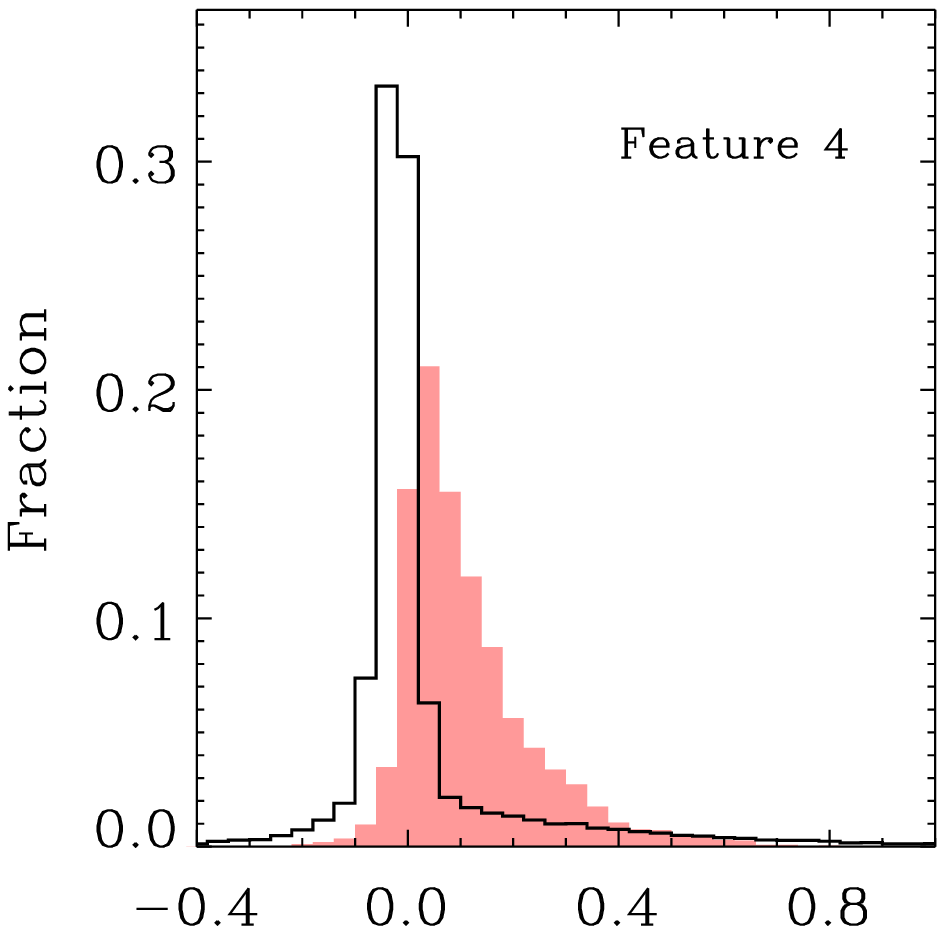}
\includegraphics[width=0.3\linewidth]{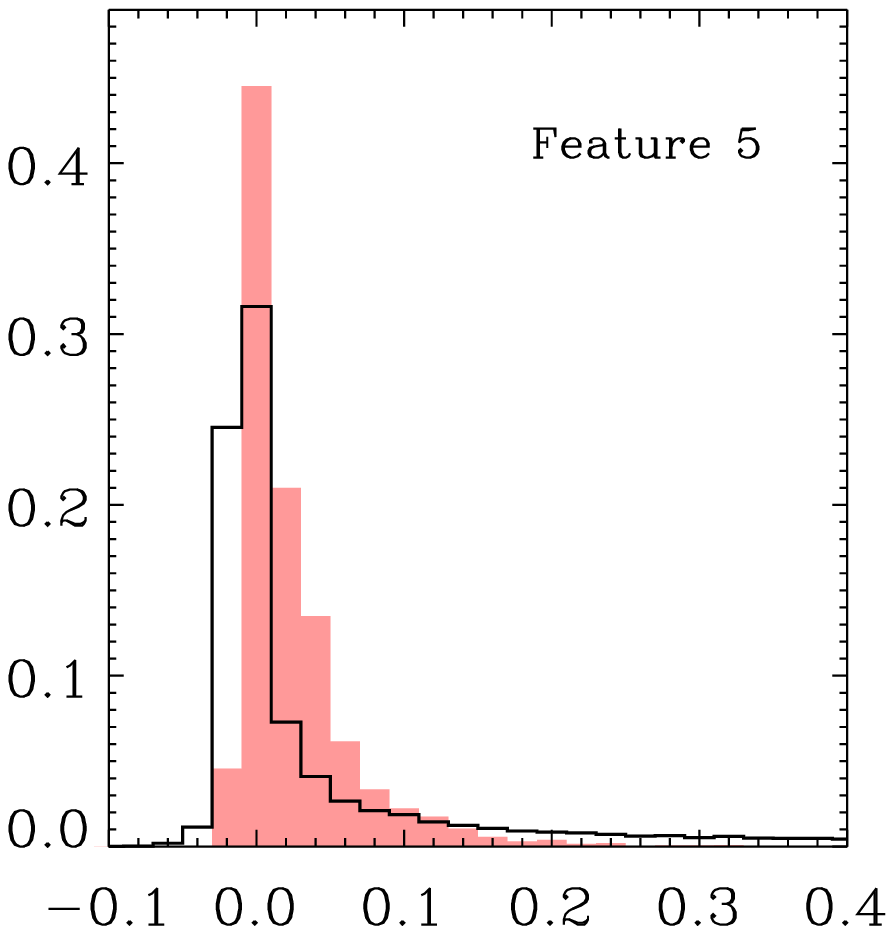} \\
\caption{Feature distributions of two classes in the training data set. \label{fig:featuredist}}
\end{figure}

All these features were extracted from each pixel of both pre-labeled training set and unlabeled test set. The distributions of the features in the training set are shown in Figure~\ref{fig:featuredist}. These features show different distributions and scales between classes. However, the different scales in the feature space may mislead the classifier to pay too much attention on certain feature. To avoid this situation, we normalized the features according to the scales of LHW class to make sure the features are equally weighted. 

\subsection{Optimizing with Cross-validation} \label{sec:optimize}

The performance of trained SVM model mainly depends on two free parameters: trade-off factor $C$ and kernel parameter $\gamma$. \citet{bea11} illustrated the impacts of these parameters on the boundary in a 2-dimensional feature space in detail. In our cases, $C$ determines trade-off between LHW and non-LHW class when misclassification occurs. A model trained with high $C$ is more likely to find every pixel of LHW class but meanwhile bring in more pixels of non-LHW class because a LHW pixel misclassification introduces higher penalty to the model than a non-LHW one. According to the study of \citet{mor99}, the trade-off factor of imbalanced classes could be chosen near the ratio between sample number of the non-LHW and LHW classes. After excluding the non-LHW pixels that touch the edge of feature space, we derived a ratio around 1.5, and tested three factors in our training process: 1.0, 1.5, and 2.0. The other parameter, $\gamma$, controls the curvature of decision boundaries. A model trained with higher $\gamma$ focuses on small scale in feature space, and gives complicated boundaries to separate classes as many as possible even if some of them are noise. As our feature space was normalized, we tested $\gamma$ of different magnitudes ranging from 0.01 to 100.


Training models with cross-validation can provide appropriate parameters to avoid underfitting and overfitting which are common issues in machine learning. A small portion of the training set ($\sim$38\%) is used as validation set (dashed line box in Figure~\ref{fig:guidemap}) in this controlled experiment, while the rest is regarded as training set. While traversing all parameter combinations, we trained models using the training set. The models were then applied to check whether LHWs could be retrieved in training set and validation set. When underfitting occurs, a model that fails to catch the trend in the training set performs almost equally bad in validation set. On the other end, an overfit model performs well in the training set by capturing the noise in it, which is not able to generalize to the validation set. By comparing all models, we eventually chose a relatively loose model with $C=1.0$ and $\gamma=10$ since it performed similar in both sets, recovered all the LHWs in the validation set, and included acceptable quantity of contaminations.

\begin{figure}[htbp]
\centering
\includegraphics[width=0.8\linewidth]{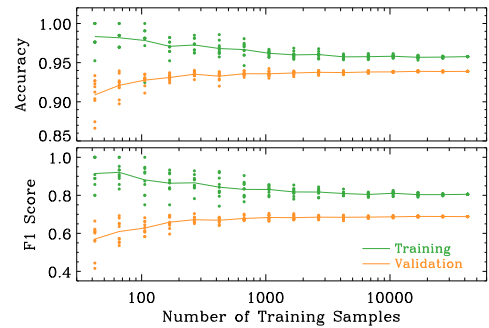}\\
\caption{Learning curves expressed in accuracy and F1 score. Each dot represents the performance of a model in the training set (green) or validation set (orange). \label{fig:learningcurves}}
\end{figure}

Learning curves are widely used to diagnose the generalization ability of models from a training set incrementally. Figure~\ref{fig:learningcurves} shows a form of learning curves in which we use accuracy and F1 score (for more details see Appendix~\ref{appendix}) to evaluate the performance of a model in certain set.
As an increasing number of random pixels is used for training, the two curves converge toward a relatively high accuracy and F1 score which suggests neither underfitting nor overfitting is significant in our model. Furthermore, though our model may be improved by collecting more training data, such approach is likely to be inefficient. The robustness of our model could also be tested by repeating the cross-validation procedure using incomplete training set. The feature distributions were redefined according to data used for training. Ten experiments, in which 37\% of the date are randomly removed from the training set, show that on average the same number of LHW candidates could be reproduced as using a full set. $\sim$93\% of the candidates overlap, which include all LHWs found by eye.



\subsection{Prediction and Grading} \label{sec:predict}

The trained classifier was applied to predict all the pixels in our mapping region, including the training region. Since each pixel was calculated independently, a large number of isolated pixels were accidentally classified as LHW. We prepared a filter to only retain the contiguous pixels occupying more than 3 pixels in the spatial dimensions ($\sim$beam size) and spanning longer than 12 velocity channels ($\sim2$~K~km~s$^{-1}$). These pixels were labeled and grouped into 960 LHW candidates in the region.

Then we followed the method mentioned in section~\ref{sec:train} to examine each candidate carefully. In total, 2880 diagrams were checked. As a result, 127 candidates fulfilled all three criteria and were classified as grade A. 290 candidates were classified as grade B, which satisfy the first two criteria but either shows shorter velocity shift ($1<\Delta v<2$~km~s$^{-1}$) or has other nearby components with the same velocity in the p-v maps. The remaining 543 candidates which are basically contaminations in our searching process were classified as grade C and discarded. Blueshifted and redshifted LHWs were then paired to constitute bipolar candidates according to their positions and velocities. The Paired LHWs should be close to a gas condensation and extend from the same velocity component of $^{13}$CO. To derive a reliable outflow catalog from the candidates, we only kept the bipolar candidates, and mono-polar ones of grade A, which will result in a higher bipolar ratio in our catalog. Mono-polar candidates in the Outer arm were further removed. Eventually, we identify 130 outflow candidates.


\section{Result} \label{sec:result}

\subsection{List of Outflows}

\begin{figure*}
\centering
\includegraphics[width=0.25\linewidth]{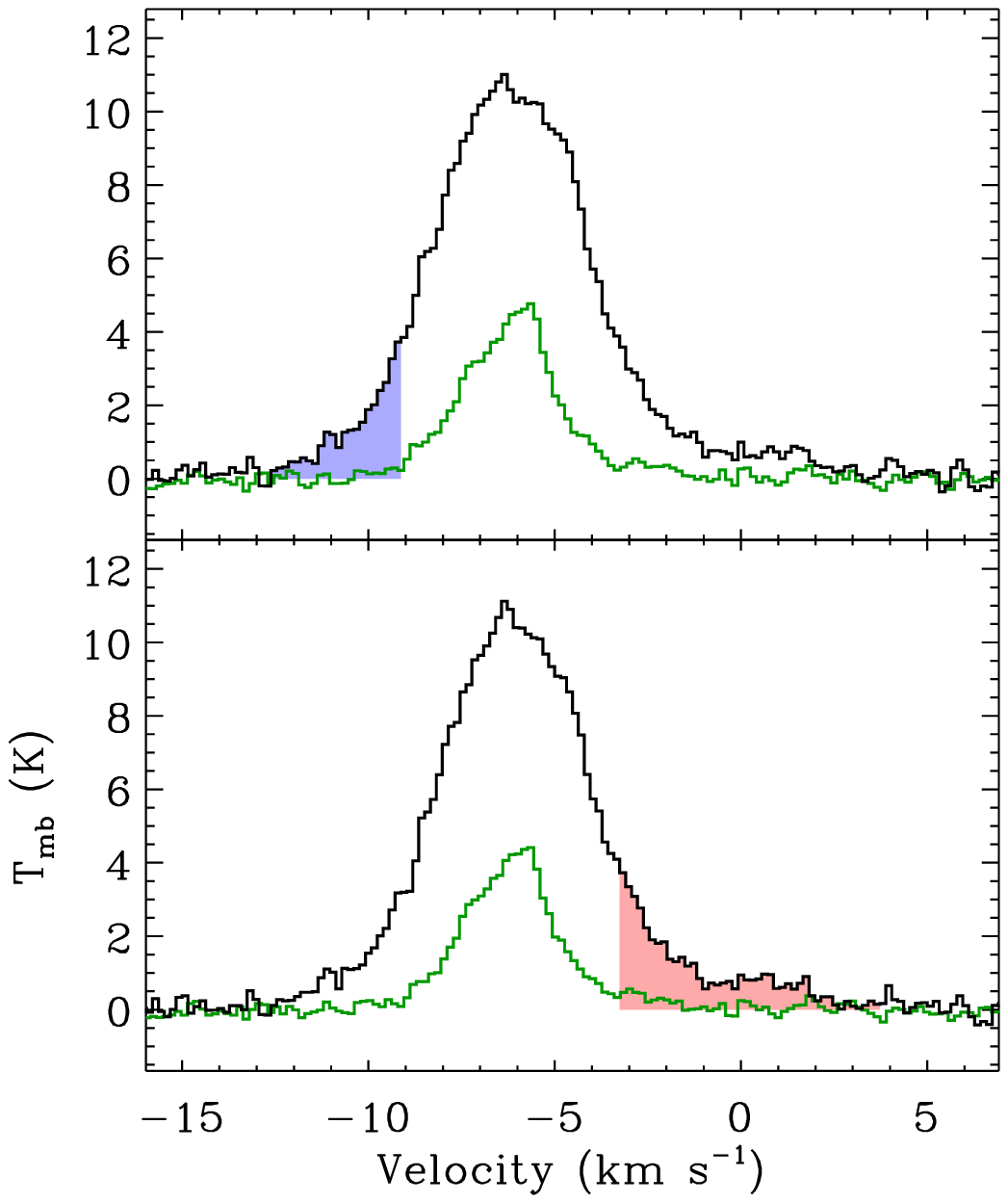}
\includegraphics[width=0.285\linewidth]{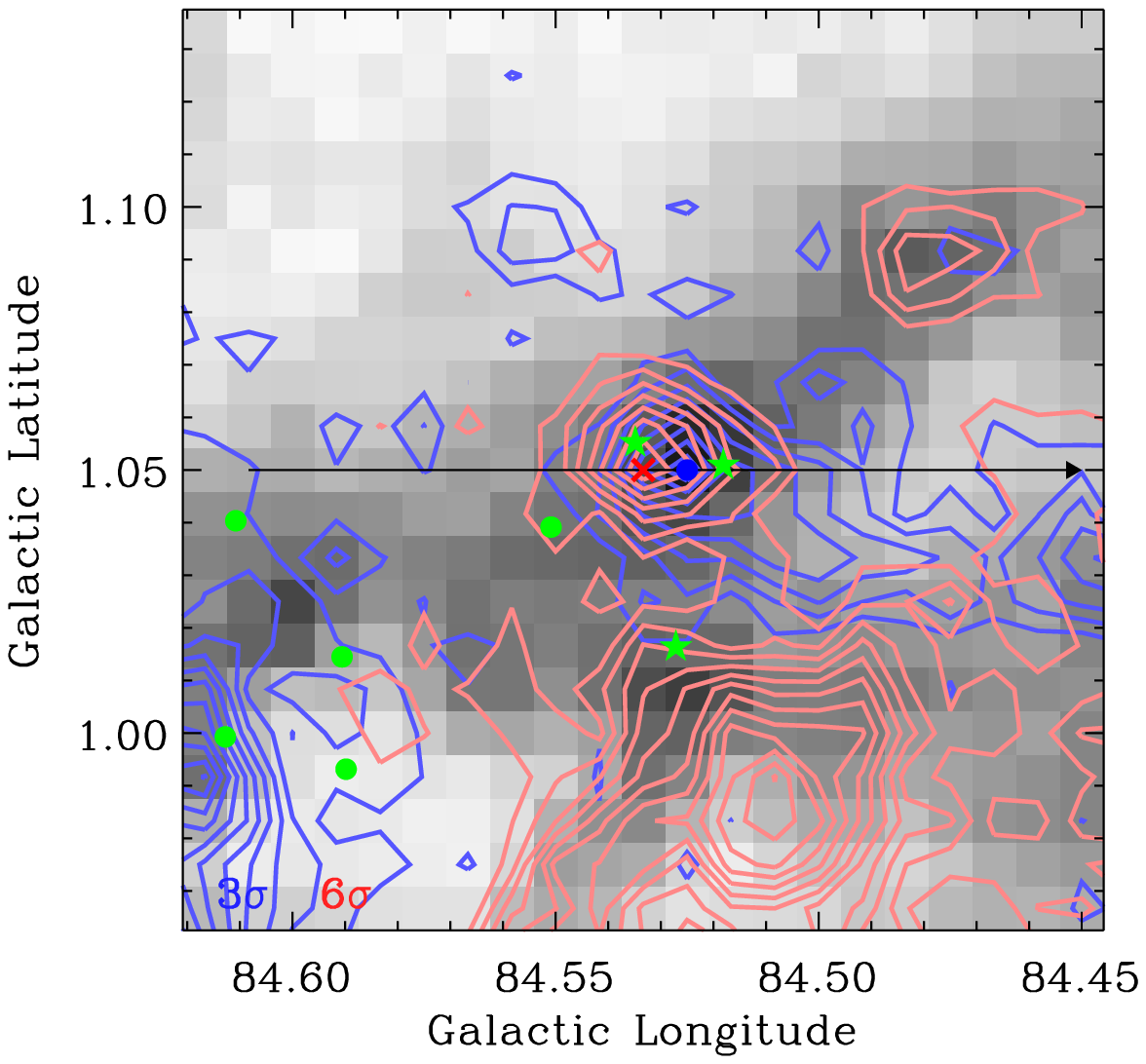}
\includegraphics[width=0.36\linewidth]{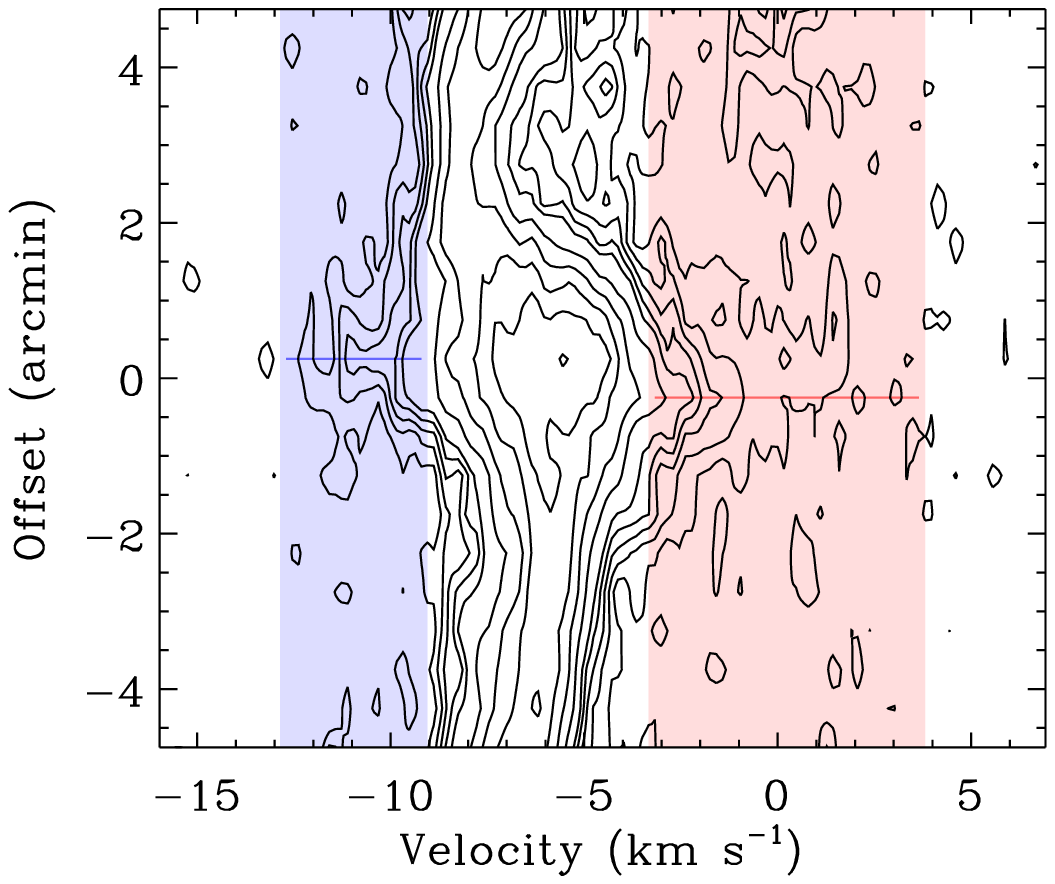}\\
\includegraphics[width=0.25\linewidth]{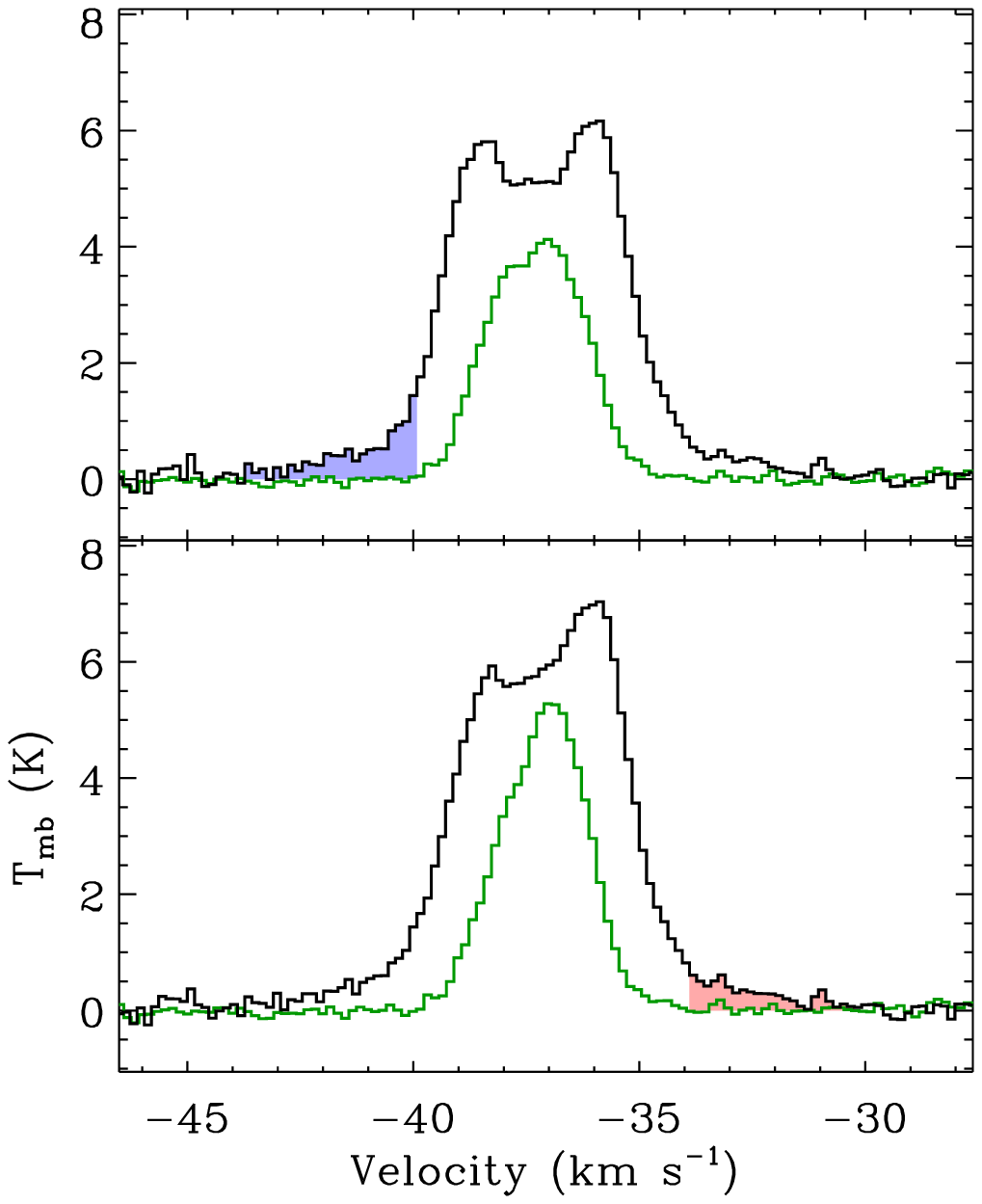}
\includegraphics[width=0.285\linewidth]{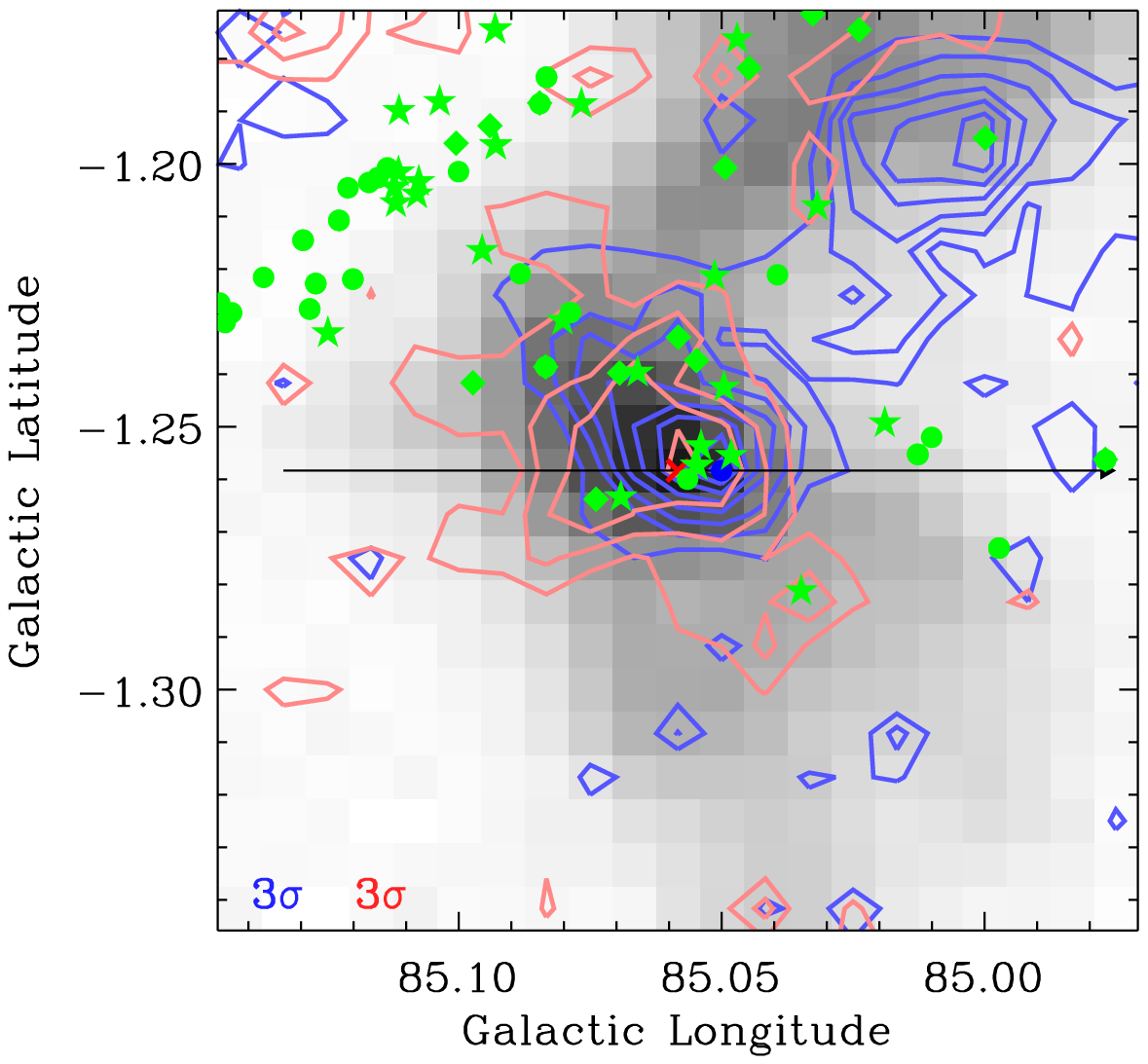}
\includegraphics[width=0.36\linewidth]{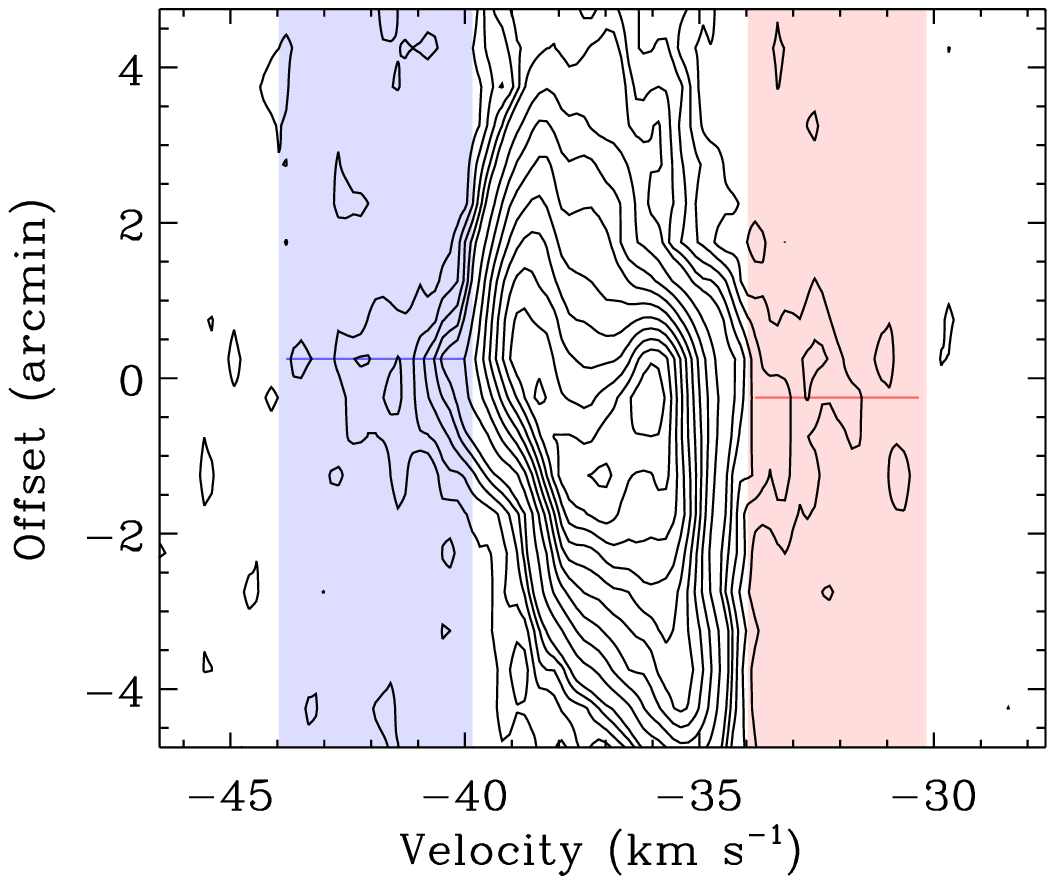}\\
\caption{Spectra (left), integrated intensity map (middle), and P-V diagram (right) of DMOC-0016 (top row), 0058 (bottom row) as examples of the new detected outflows. The red `x' and blue dot on the integrated intensity map mark the positions where the spectra are extracted. The black long arrow indicates the axis of the P-V map. The contours are at intervals of $2sigma$ starting from the levels marked in the lower left corner. The green stars, diamonds, and dots on the map represent YSOs of Class I, flat, and II, respectively. \label{fig:outflowmap}\\
(The complete figure set (390 images) is available in the online journal.)}
\end{figure*}

Among the 130 outflow candidates, 77 (59.2\%) samples show bipolar structure; 29 (22.3\%) and 24 (18.5\%) samples are mono-blue and mono-red polar, respectively. In Table~\ref{tab:outflow}, we list all outflows detected in this work with brief comments. Each outflow is referred as a ``Delingha Molecular Outflow Candidate'' (DMOC). The position and velocity of blueshifted and redshifted lobes are cataloged separately with the coordinates representing the peak positions of the lobes. Some bipolar outflows contain more than two lobes which indicates there are multiply peaks in its blueshifted or redshifted lobe. The velocity range and grade of the outflows are the direct results of the outflow searching procedure. Since the studies towards the molecular clouds in our mapping region are limited, 118 outflows are detected using molecular spectra line for the first time among all detections. We also detect the counterparts of two known mono-polar outflows, and classify them as bipolar outflows in our catalog. The spectra, integrated intensity map, and P-V diagram of two new detected outflows are shown in Figure~\ref{fig:outflowmap} as examples.

\begin{deluxetable*}{ccccr@{, }lccccl}
\addtolength{\tabcolsep}{-1pt}
\tablecaption{List of detected outflows \label{tab:outflow}}
\tablewidth{0pt}
\tablehead{
\colhead{Name} &
\colhead{Lobe} &
\colhead{$l$} &
\colhead{$b$} &
\multicolumn{2}{c}{Velocity range} &
\colhead{$v_{\rm sys}$} &
\colhead{Grade} &
\colhead{New} &
\colhead{YSO} &
\colhead{Comments}
\\
\colhead{} &
\colhead{} &
\colhead{(\arcdeg)} &
\colhead{(\arcdeg)} &
\multicolumn{2}{c}{(km~s$^{-1}$)} &
\colhead{(km~s$^{-1}$)} &
\colhead{} &
\colhead{detection} &
\colhead{class} &
\colhead{}
}
\startdata
DMOC-0001 & Red  & 83.767 & $+0.125$ &( $  2.7$ & $  6.5$ ) & 0.3 & A & Y & I/II\\
DMOC-0002 & Blue & 83.850 & $-2.092$ &( $ -5.1$ & $ -2.9$ ) & 1.7 & B & Y & II\\
         & Red  & 83.850 & $-2.100$ &( $  6.2$ & $  9.5$ )& 1.7& B\\
DMOC-0003 & Red  & 83.875 & $+0.092$ &( $  3.3$ & $  7.5$ )& 1.0 & A & Y & -\\
DMOC-0004 & Red  & 83.892 & $-0.892$ &( $  7.9$ & $  9.8$ )& 5.1 & A & Y & -\\
DMOC-0005 & Blue & 83.942 & $+0.783$ &( $-15.1$ & $ -9.4$ )& $-$5.4 & A & Y & -         & 3C 423\\
         & Red  & 83.925 & $+0.792$ &( $ -1.6$ & $  5.4$ )& $-$5.4 & A\\
DMOC-0006 & Blue & 83.967 & $+0.033$ &( $ -5.1$ & $ -1.9$ )& 0.6 & A & Y & I/flat/II & IRAS 20472+4338\\
         & Red  & 83.967 & $+0.050$ &( $  3.2$ & $  6.2$ )& 0.6 & A\\
DMOC-0007 & Blue & 84.017 & $+0.008$ &( $ -5.1$ & $ -1.4$ )& 0.5 & B & Y & I/II\\
         & Red  & 84.008 & $+0.008$ &( $  3.5$ & $  6.2$ )& 0.5 & B\\
DMOC-0008 & Blue & 84.108 & $-0.550$ &( $ -7.5$ & $ -2.7$ )& 1.3 & A & Y & -\\
DMOC-0009 & Blue & 84.167 & $-1.733$ &( $ -3.5$ & $ -0.2$ )& 2.1 & A & Y & -\\
DMOC-0010 & Blue & 84.292 & $+0.900$ &( $-10.3$ & $ -8.6$ )& $-$3.8 & A & Y & -         & IRAS 20444+4425\\
\enddata
\tablecomments{List of the identified molecular outflow candidates in this work. Columns are outflow name, lobe, position (Galactic longitude \& latitude), velocity range, systemic velocity, grade assigned in searching procedure, new detected outflow in this work or not, class of associated YSOs, and comments.\\
REFERENCE: (1)\citealt{arm11}; (2)\citealt{bal03}; (3)\citealt{bal14}; (4)\citealt{dun12}; (5)\citealt{Li16}; (6)\citealt{sun07}; (7)\citealt{tou11}; (8)\citealt{urq11}; (9)\citealt{wu04}.\\
(This table is available in its entirety in machine-readable and Virtual Observatory (VO) forms in the online journal. A portion is shown here for guidance regarding its form and content.)}.
\end{deluxetable*}

\figsetstart
\figsetnum{5}
\figsettitle{Maps for all outflow samples}

\figsetgrpstart
\figsetgrpnum{4.16}
\figsetgrptitle{DMOC_0016}
\figsetplot{fig04set/0016_spec.eps}
\figsetplot{fig04set/0016_map.eps}
\figsetplot{fig04set/0016_pv.eps}
\figsetgrpnote{Spectra, integrated intensity map, and P-V diagram of DMOC-0016}
\figsetgrpend

\figsetgrpstart
\figsetgrpnum{4.58}
\figsetgrptitle{DMOC_0058}
\figsetplot{fig04set/0058_spec.eps}
\figsetplot{fig04set/0058_map.eps}
\figsetplot{fig04set/0058_pv.eps}
\figsetgrpnote{Spectra, integrated intensity map, and P-V diagram of DMOC-0058}
\figsetgrpend

\figsetend

In Figure~\ref{fig:guidemap}, we map the locations of the identified outflows in each velocity component, where the blueshifted and redshifted lobes are color-coded symbols. Most outflows are concentrated in the clouds with high column density. Outflows mainly cluster in the Gulf of Mexico (L935), Pelican's Head (L933, L936), IC~5146 (L1040/1042/1055), G084.9$-$0.4, and Sh-2~124 region. Groups of outflows arrange on the filamentary molecular cloud, such as L931, IC~5146, G084.9$-$0.4. There are 97, 31, and 2 outflows identified in the Local, Perseus, and Outer arms, respectively. 

As potential driving source of outflows, YSOs are found to be associated with some of our molecular outflows. We list the class of YSOs around each outflow within 3\arcmin\ in the 9th column of Table~\ref{tab:outflow} where two YSO catalogs are included. We identified YSOs in our mapping region following the detection and classification scheme given by \citet{koe14}. Such scheme identifies Class I and II stars based on near- and mid-infrared colors of WISE and 2MASS photometry. We further included the YSO catalog identified by \citet{reb11} in the North American and Pelican Nebulae region for complement. 61 of our samples ($\sim46.9\%$) are associated with Class I/flat YSOs, while 82 outflows ($\sim63.1\%$) are located near Class II YSOs. Some outflows are located close to several YSOs which may be of different classes. Observations with higher resolution are required to further distinguish the driving sources. Furthermore, 36 ($\sim27.7\%$) outflows have no YSO associated, half of which locate in the Local arm. It is evident that the sensitivities of infrared observations limit the detection of distant or deeply embedded YSOs as outflow driving sources. We should also consider the possibility that some of the outflows are driven by protostars at an earlier stage than Class I. As pointed out by \citet{mot17}, Class 0 objects tend to present a higher outflow mass and momentum injection rate than those of Class I.


We list the associated infrared sources and outflow indicators from other band in the last column of Table~\ref{tab:outflow}. 29 samples associate with IRAS point sources. Different types of shock indicators such as Herbig–Haro (HH) objects, molecular hydrogen emission-line objects (MHOs), and water masers are found in the lobes, which improves the reliability of our detections \citep{arm11, bal03, bal14, sun07, tou11, urq11}.

\subsection{Physical Properties}\label{sec:properties}

It is important to derive accurate physical properties such as mass, momentum, and mechanical energy of the outflow, since they will help us to understand the feedback activities of forming stars to their immediate environment. Prior to the calculation, the distance of the outflows need to be settled. We use the distance estimator given by \citet{rei16} which provides a combined result of parallax and kinematic distance using a Bayesian approach. The position and velocity of $^{13}$CO for each outflow are extracted and taken as input parameters, and the output distance with the highest probability is designated. The calculated distance of Sh-2~124 agrees with that reported by \citet{fos15}. However, the calculated accuracy is limited for the nearby clouds with velocity around 0~km~s$^{-1}$, and such method fails to converge at relative high latitude such as in IC~5146, L1042, and L1045. \citet{gre15} provides a method based on Pan-STARRS 1 and 2MASS photometry to derive more accurate distances for these nearby clouds. We choose the smaller distance moduli if a cloud is associated with an optical or near-infrared dark cloud. Such method gives a distance of 0.9~kpc for those high latitude clouds, which is similar to the distance of 0.95~kpc given by \citet{har08}. However, \citet{her02} and \citet{nun16} suggested a larger measurement of 1.1-1.4~kpc, which will multiply the calculated mass by a factor of 1.3-2.2.

We list the properties of the outflows in Table~\ref{tab:property}, in which the $v_{\rm avg}$ column presents the average velocity of $^{12}$CO lobe relative to the systemic velocity of corresponding $^{13}$CO emission. The average velocity ranges from 2.1 to 10.5~km~s$^{-1}$. The size of a lobe is given by measuring the extent along Galactic longitude and latitude of the contour at half-maximum, while its length is measured from the 3$\sigma$ contour to the nearest $^{13}$CO clump or YSO. While measuring the scale and velocity of the outflows, their inclinations are an intractable problem, on which several correction methods have been proposed in the literature \citep{dun14}. The proper motion studies of HH objects provide the most reliable measurement of the inclination. However, such correction is impossible for our sample, since only a few lobes associate with HH object and no proper motion study has been done. Therefore no inclination correction is applied to the parameters listed in Table~\ref{tab:property}. If a mean inclination of 57.3\arcdeg\ is adopted (assuming all orientations are uniformly distributed in the space), the average velocity and length will be scaled up by a factor of 1.85 and 1.19, respectively.

We further calculate the column density, mass, momentum, and energy of each lobe under the assumption of local thermodynamic equilibrium (LTE) by following the standard method given by \citet{wil09} and \citet{cur10}. Without opacity correction, the column density can be calculated with
\begin{displaymath}
\left( {N({\rm ^{12}CO}) \over {\rm cm}^{-2}} \right) = 4.17 \times 10^{13} {T_{\rm ex} / {\rm K} \over {\rm exp}(-5.53 {\rm K} / T_{\rm ex})} \left( {\int T_{\rm mb}(v) {\rm d}v \over {\rm K\ km\ s}^{-1}} \right),
\end{displaymath}
after which a CO abundance of 10$^{-4}$ \citep{fre82} and a mean molecular weight of 2.72 \citep{bru10} is adopted to derive the column density of molecular hydrogen. The column densities in each velocity slice are then summed over each lobe to derive its mass, momentum, and energy which are defined as
\begin{displaymath}
M = \int m(v) {\rm d}v;\\
\end{displaymath}
\begin{displaymath}
P = \int m(v)|v-v_0| {\rm d}v;\\
\end{displaymath}
\begin{displaymath}
E = {1 \over 2} \int m(v)|v-v_0|^2 {\rm d}v.
\end{displaymath}
For most of our candidates, the $^{12}$CO line wing emissions are assumed to be optically thin, since no $^{13}$CO line wing is detected within the velocity range of $^{12}$CO lobes except for a few ones. There are weak $^{13}$CO line wing features detected in the blueshifted lobe of DMOC-0025, 072, 123, 128 and the red-shifted lobe of DMOC-0016, 072. Their optical depths could be estimated following the relation given by \citet{hat99}
\begin{displaymath}
{T_{\rm A}^*({\rm ^{12} CO}) \over T_{\rm A}^*({\rm ^{13} CO})} = \left( {\nu_{12} \over \nu_{13}} \right)^2 {[{\rm ^{12}CO}] \over [{\rm ^{13}CO}]} {1-e^{-\tau_{12}} \over \tau_{12}},
\end{displaymath}
where [$^{12}$CO]/[$^{13}$CO] = 62, is the relative abundance between the two species \citep{lan93}. The resulting optical depths range from 6 to 10, which will bring a correction factor of $\tau/(1-e^{-\tau})$ to the parameters concerning mass. In the above calculations, we use an uniform excitation temperature of 25~K, which could be differ from the real situation. A large range of excitation temperature has been discussed by several studies such as \citep{hir01,sto06}. Assuming a lower temperature of 10~K as \citet{dow07} will decrease the mass, momentum, and energy by a factor of 1.8, while a higher value of 50-100~K as \citet{hir01} will increase these parameters by a factor of 1.8-3.4. It is also noteworthy that outflow emission at low velocity which may mix with ambient gas are omitted. At least a factor of two will be introduced to take such correction into consideration \citep{mar85}. \citet{dun14} discussed the correction factors for the parameters in detail based on line ratio analysis to low-J CO outflows and gave a higher mean factor of 7. As a result, the calculated mass, momentum, and energy are lower limits.

\begin{deluxetable*}{ccccccccccc}
\addtolength{\tabcolsep}{-1pt}
\tablecaption{Physical properties of outflows \label{tab:property}}
\tablecolumns{11}
\tablewidth{0pt}
\tablehead{
\colhead{Name} &
\colhead{Lobe} &
\colhead{Distance} &
\colhead{$v_{\rm avg}$} &
\colhead{Size} &
\colhead{Mass} &
\colhead{Momentum} &
\colhead{Energy} &
\colhead{Length} &
\colhead{$t_{\rm dyn}$} &
\colhead{$L_{\rm flow}$}
\\
\colhead{} &
\colhead{} &
\colhead{(kpc)} &
\colhead{(km s$^{-1}$)} &
\colhead{(\arcmin)} &
\colhead{(M$_\odot$)} &
\colhead{(M$_\odot$ km s$^{-1}$)} &
\colhead{(10$^{43}$ erg)} &
\colhead{(\arcmin)} &
\colhead{(10$^5$ yr)} &
\colhead{(10$^{30}$ erg s$^{-1}$)}
}
\startdata
DMOC-0001 & Red  & 0.71$\pm$0.08 & 4.32 & 2.4$\times$2.3 &  0.4 &    1.5 &   6.1 & 3.0 &  1.4 &  0.1\\
DMOC-0002 & Blue & 1.42$\pm$0.16 & 5.67 & 2.9$\times$2.9 &  3.3 &   18.3 & 103.1 & 2.1 &  1.5 &  2.2\\
         & Red  & 1.42$\pm$0.16 & 6.16 & 3.2$\times$2.9 &  4.8 &   26.6 & 150.8 & 1.5 &  1.0 &  4.9\\
DMOC-0003 & Red  & 0.71$\pm$0.08 & 4.42 & 4.0$\times$2.8 &  0.8 &    3.2 &  14.0 & 4.0 &  1.8 &  0.2\\
DMOC-0004 & Red  & 1.49$\pm$0.20 & 3.80 & 2.5$\times$2.4 &  1.3 &    4.7 &  17.0 & 1.2 &  1.3 &  0.4\\
DMOC-0005 & Blue & 2.76$\pm$0.61 & 6.80 & 2.5$\times$1.7 & 17.2 &  103.7 & 668.2 & 3.1 &  3.6 &  5.9\\
         & Red  & 2.76$\pm$0.61 & 7.33 & 2.4$\times$1.7 & 17.9 &  115.9 & 826.1 & 1.4 &  1.5 & 17.4\\
DMOC-0006 & Blue & 0.90$\pm$0.10 & 4.06 & 2.2$\times$2.6 &  0.7 &    2.6 &  10.0 & 2.1 &  1.3 &  0.2\\
         & Red  & 0.90$\pm$0.10 & 4.11 & 4.1$\times$2.9 &  1.4 &    5.2 &  19.7 & 2.5 &  1.6 &  0.4\\
DMOC-0007 & Blue & 0.90$\pm$0.10 & 3.78 & 2.0$\times$1.5 &  0.7 &    2.3 &   8.3 & 2.8 &  1.9 &  0.1\\
         & Red  & 0.90$\pm$0.10 & 4.31 & 2.9$\times$3.3 &  1.1 &    4.4 &  17.6 & 2.4 &  1.4 &  0.4\\
DMOC-0008 & Blue & 0.71$\pm$0.08 & 6.35 & 2.9$\times$2.6 &  0.9 &    5.2 &  30.8 & 3.6 &  1.1 &  0.9\\
DMOC-0009 & Blue & 1.52$\pm$0.19 & 3.96 & 2.4$\times$2.4 &  2.4 &    8.4 &  31.7 & 3.2 &  3.5 &  0.3\\
DMOC-0010 & Blue & 2.62$\pm$0.56 & 5.61 & 1.4$\times$1.4 &  1.8 &    9.8 &  53.3 & 1.8 &  2.4 &  0.7\\
\enddata
\tablecomments{Physical properties of the identified DMOCs. Columns are outflow name, lobe, distance, average velocity away from systemic velocity, spatial size along Galactic longitude and latitude, mass, momentum, energy, length, dynamical time-scale, and outflow luminosity.\\
(This table is available in its entirety in machine-readable and Virtual Observatory (VO) forms in the online journal. A portion is shown here for guidance regarding its form and content.)}.
\end{deluxetable*}

\begin{figure*}[htbp]
\centering
\includegraphics[width=0.28\linewidth]{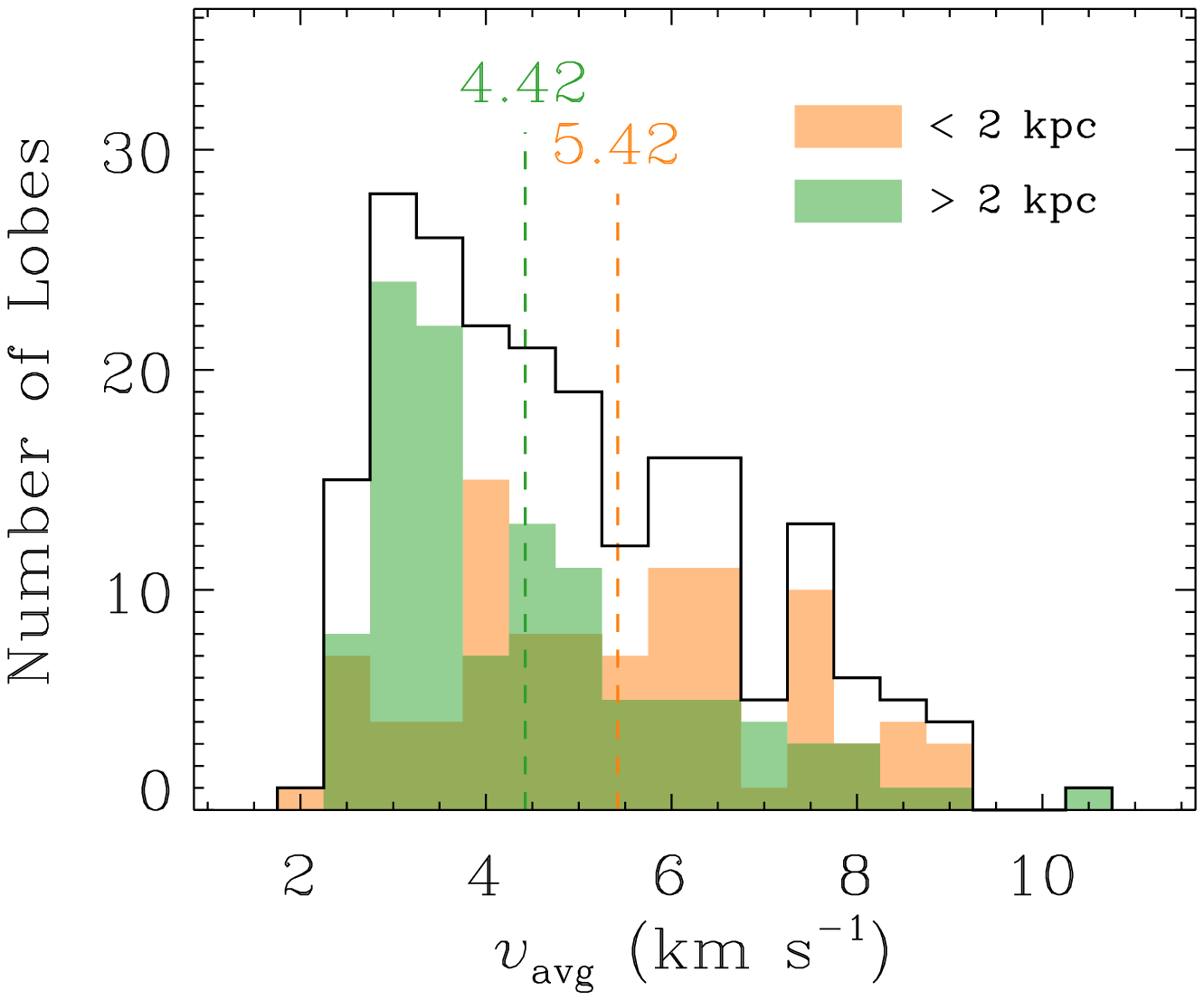}
\includegraphics[width=0.28\linewidth]{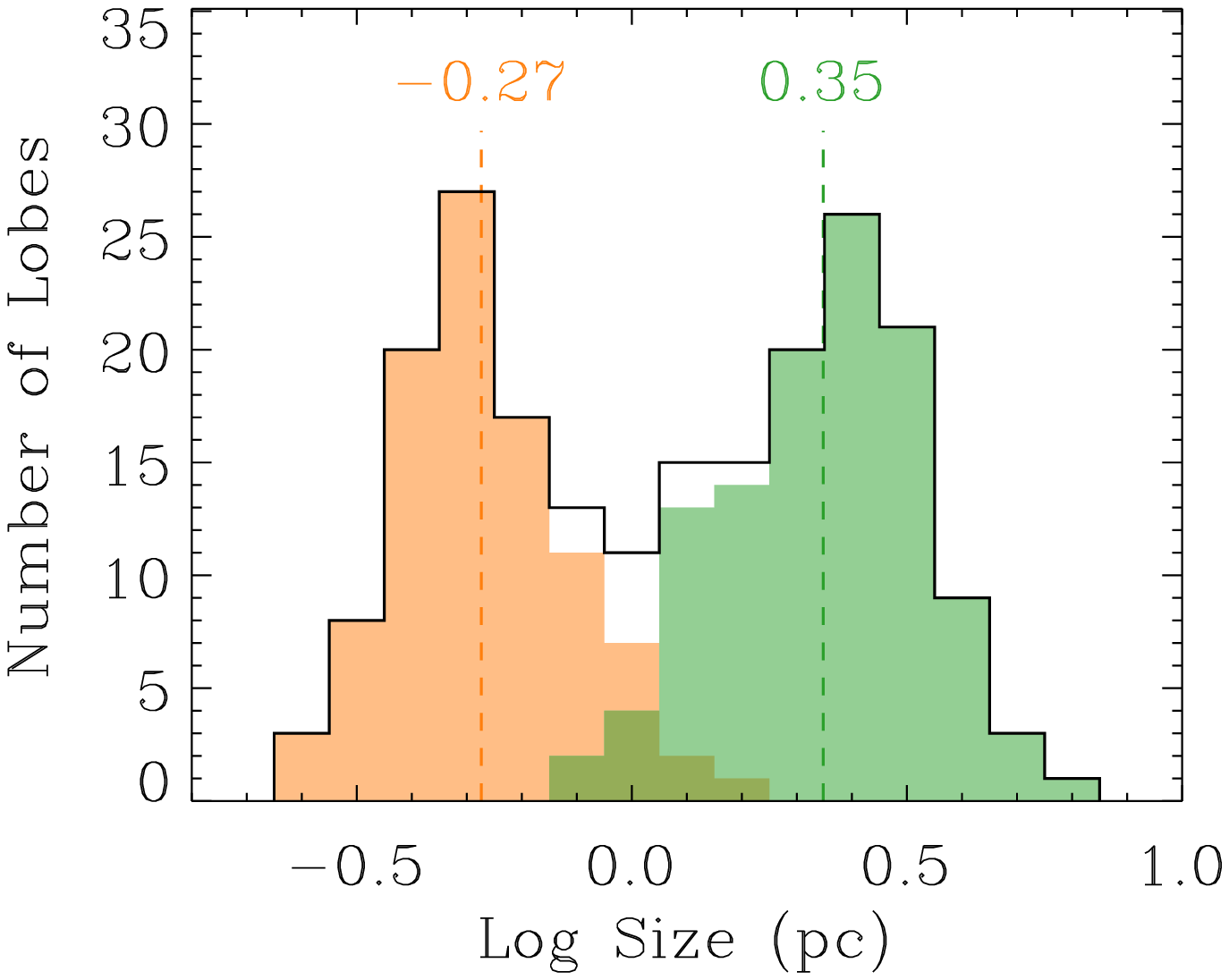}
\includegraphics[width=0.28\linewidth]{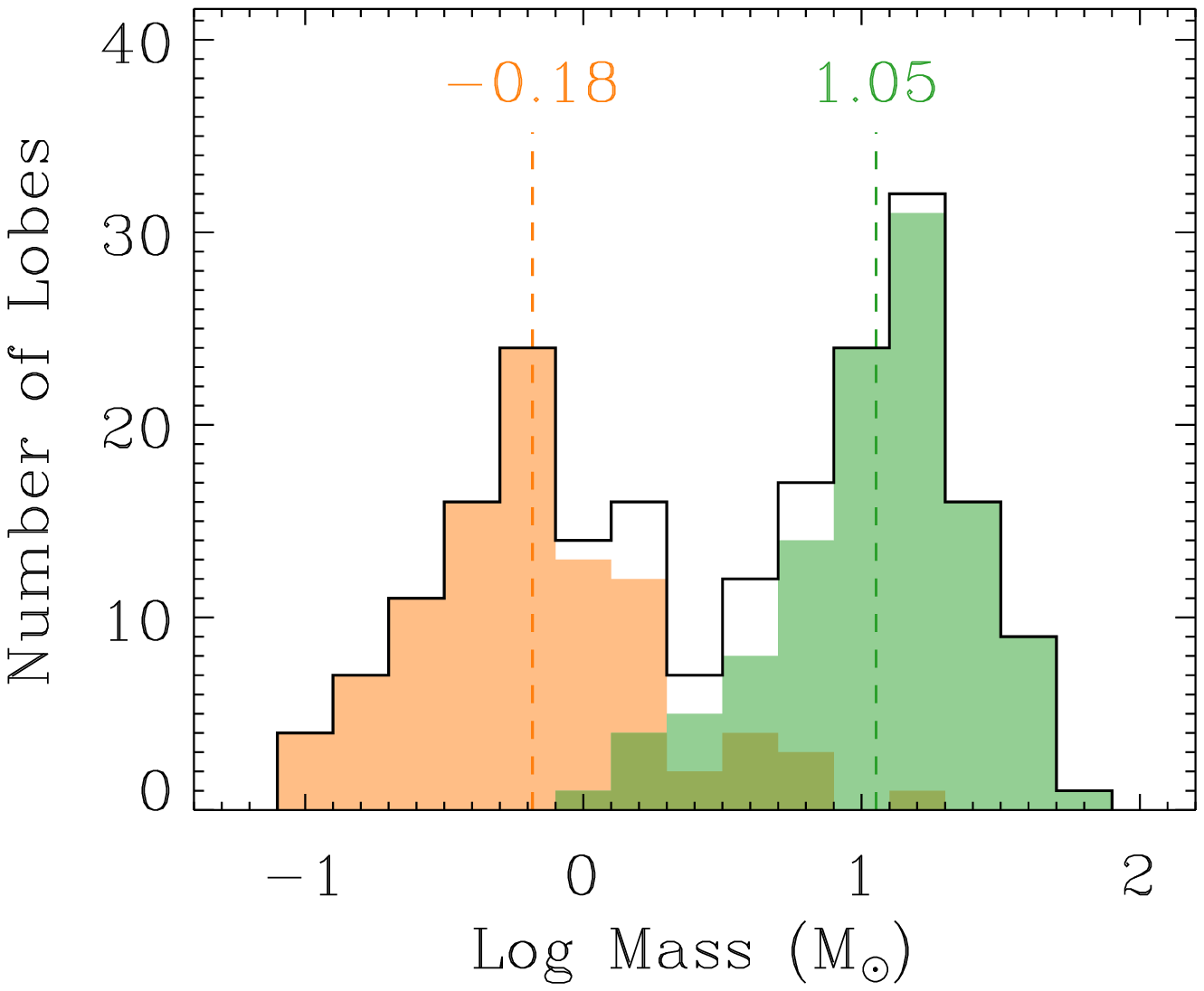}\\
\includegraphics[width=0.28\linewidth]{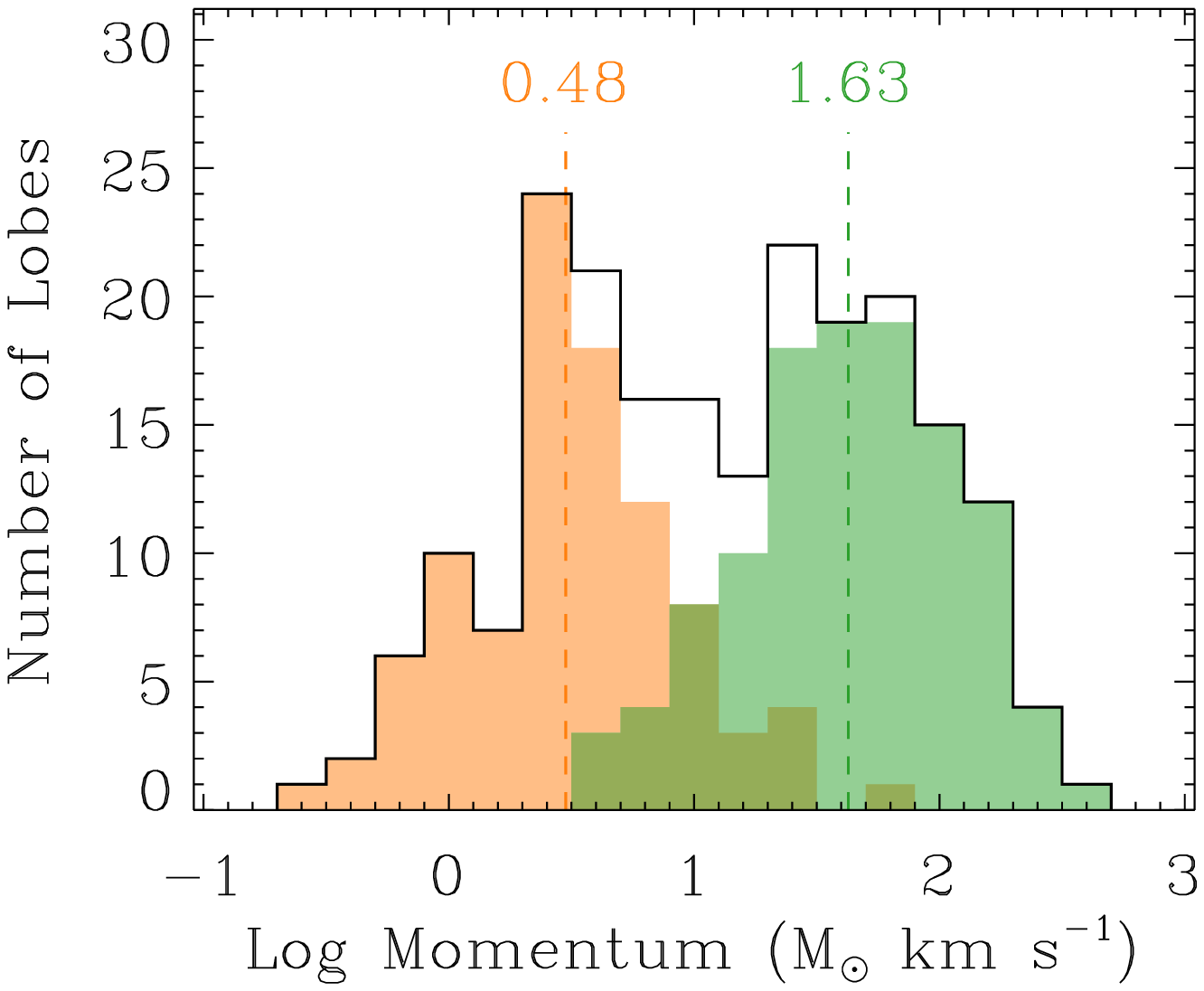}
\includegraphics[width=0.28\linewidth]{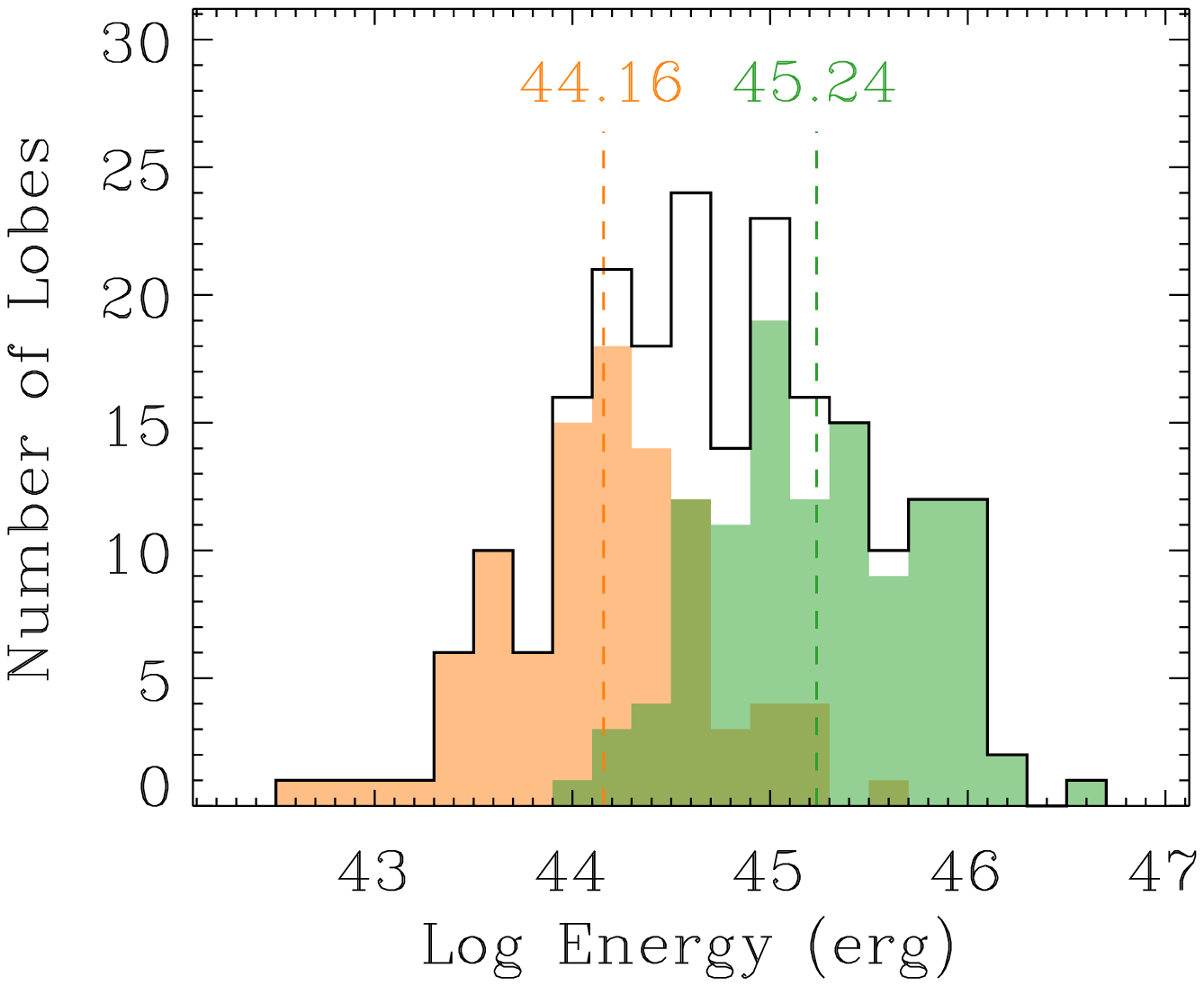}
\includegraphics[width=0.28\linewidth]{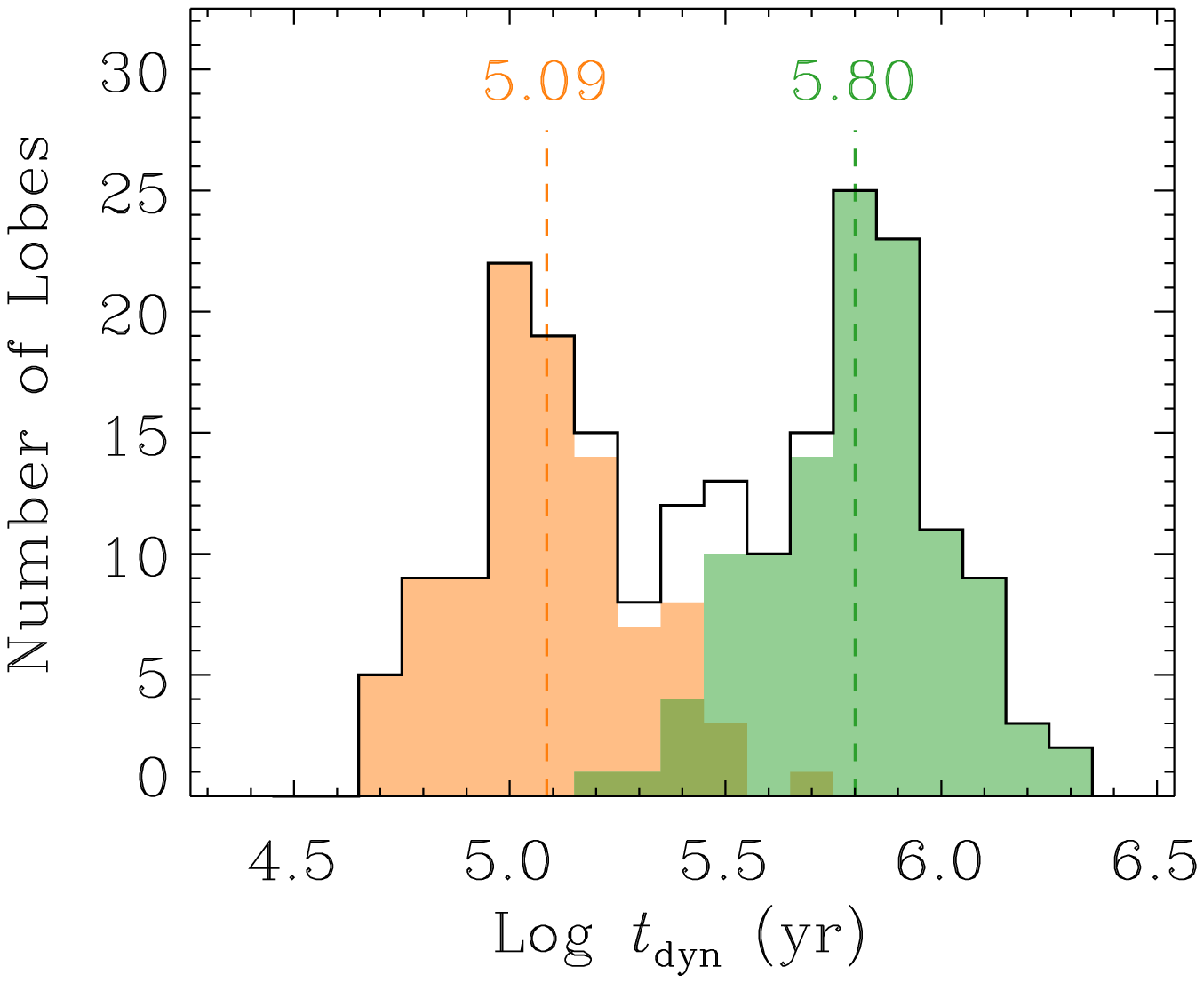} \\
\caption{Distribution of average velocity (upper left), size (upper middle), outflow mass (upper right), momentum (lower left), energy (lower middle), and dynamical time-scale (lower right). The distribution of two subgroups, less and greater than 2~kpc, are shown with orange and green bars, respectively. \label{fig:parameterdist}}
\end{figure*}

\begin{figure}[htbp]
\centering
\includegraphics[width=0.8\linewidth]{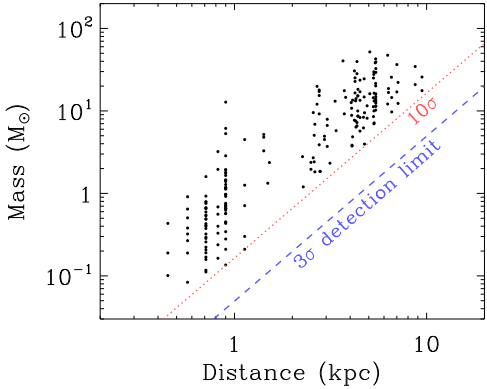}\\
\caption{Mass and distance relation. The dashed blue line indicates the $3\sigma$ detection limit calculated based on our searching procedure. The dotted red line shows the detection limit based on the identified outflows. \label{fig:distancemass}}
\end{figure}

Figure~\ref{fig:parameterdist} shows the distributions of physical properties, including average velocity, size, mass, momentum, energy, and dynamical time-scale. Double peak distributions are shown in all the diagrams with the exception of the average velocity which is measured along the line of sight and unrelated to distance. By simply dividing the samples into groups with distance greater and less than 2~kpc, we note that those peaks represent outflows at different distances and the parameters of nearby outflows seem to be lower than those of distant ones. Such divergence is illustrated more clearly by the tendencies of the lobe size and the mass varying with distance as shown in Figure~\ref{fig:distancemass}. The double peak distribution in Figure~\ref{fig:parameterdist} could be the result of uneven distribution of star forming regions along distance axis due to the spiral arms, since we only cover a short range of Galactic longitude. The trend could be explained as an observational effect that low-mass outflows fall below the detection limit in the distant molecular clouds. Furthermore, the resolution of our observations is insufficient to resolved each outflows even in some nearby clouds, which indicates that some of our candidates might be stacking of unresolved outflows from clusters or small groups of YSOs. Figure~\ref{fig:distancemass} also shows that the lower limit of mass for the identified outflows is much higher than the limit of our searching procedure, while most of the grade C candidates fall between the two limits. We could further estimate an outflow detection limit of $m{\rm (M_\odot)} > 0.16 \times D{\rm (kpc)}^2$ for the MWISP survey.

The dynamical time-scale calculated by dividing lobe length with average velocity will usually underestimate the outflow age, since the low-J CO intensity of outflow drops increasingly rapidly at high velocity \citep{dow07}. Still we could estimate the mean mass entrainment rate ($\dot{M}_{\rm out}$) and momentum injection rate or outflow force ($F_{\rm CO}$) to be $1.1\times10^{-5}$~M$_\odot$~yr$^{-1}$ and $4.7\times10^{-5}$~M$_\odot$~km~s$^{-1}$~yr$^{-1}$, repectively, which both give an uncertainty of $\pm0.5$~dex. The obtained rates represent time-averaged results over a bulk of shocked gas including a wide range of conditions, whereas the rates match the results given by \citet{mot17}.

\subsection{Outflow Clusters and Individuals}

\subsubsection{Gulf of Mexico}

\begin{figure*}[htbp]
\centering
\includegraphics[width=0.8\linewidth]{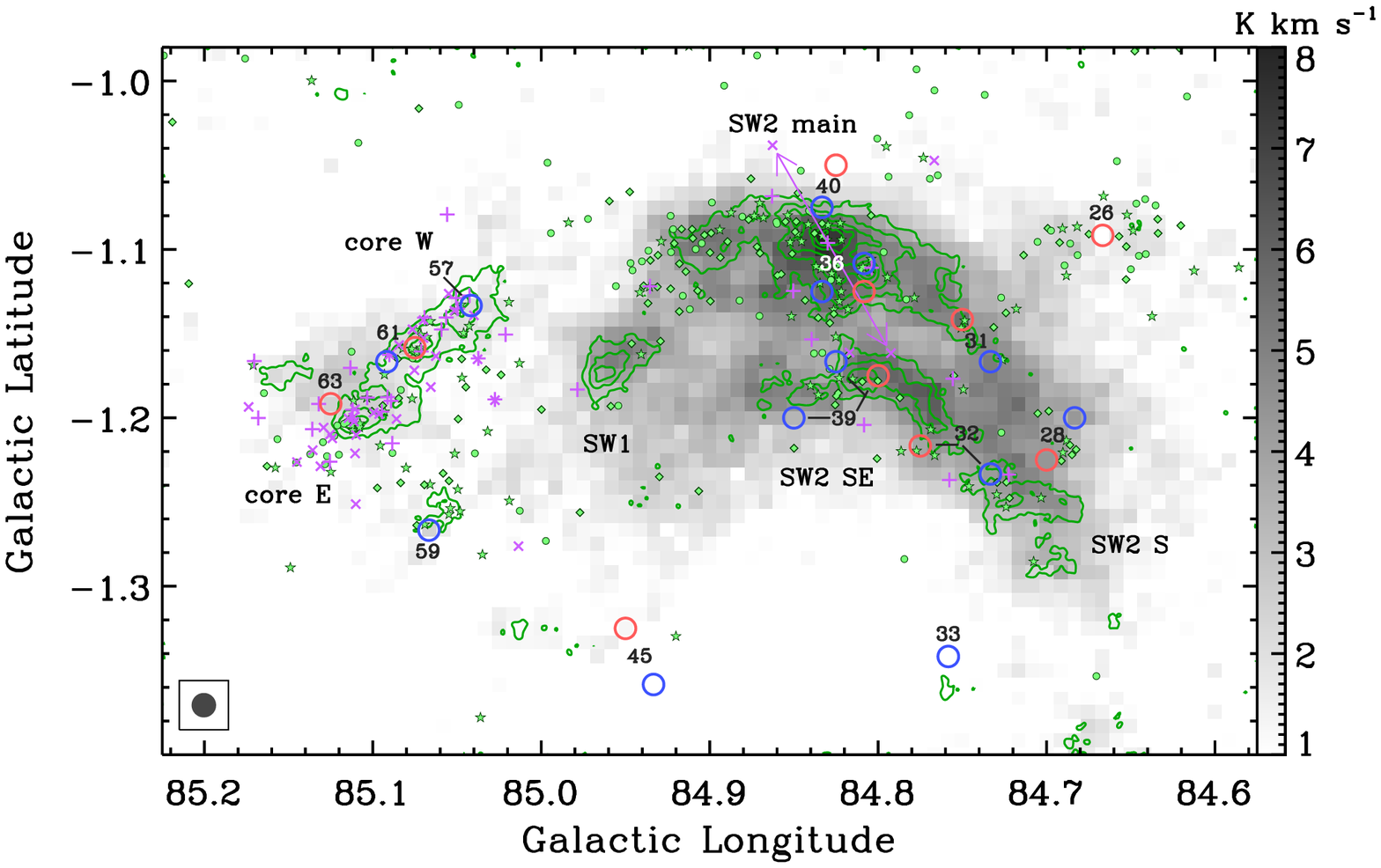}
\caption{Outflows in the Gulf of Mexico. Integrated intensity gray-scale map of C$^{18}$O is overlaid with contours showing the 1.1~mm continuum emission from the Bolocam Galactic Plane Survey (BGPS). The spectra are integrated over $-$10 to 10~km~s$^{-1}$. Blue and red circles indicate the positions of identified blue/redshifted molecular outflow lobes. Purple crosses and `x's show the locations of MHOs and HH objects. The purple double head vector indicates the size and orientation of MHO~3417, the largest outflow in this region. The green stars, diamonds, and dots represent YSOs of Class I, flat, and II, respectively. \label{fig:regionGM}}
\end{figure*}

The Gulf of Mexico region, located in the south of L935, is the most active star forming region in the North America and Pelican Nebulae. A large number of shock features such as HH objects \citep{arm11}, MHOs \citep{bal14, mak18} and their possible drivers such as Class I, II, and III YSOs \citep{gui09, reb11}, and highly embedded Class 0 YSOs have been found in this region. Together with CO outflow candidates, we show the distributions of molecular clouds and outflow features in Figure~\ref{fig:regionGM}. Just as the labels on the figure, \citet{bal14} further subdivided the region into three sub-regions, which are the elongated ``Gulf core'' in the east, the compact ``Gulf SW1'' in the middle, and the gas-rich filaments ``Gulf SW2'' in the west.

The clumps in ``SW2'' region exhibit large and strong C$^{18}$O and 1.1~mm emission as shown in Figure~\ref{fig:regionGM}, and are therefore confirmed to be massive ones \citep{zha11}. Although a great number of YSOs, as the outflow drivers, emerge in the clumps, shock features are rarely spotted in this region. Our study identified 14 molecular lobes, which show different spatial distributions from that of shocks. In the ``SW2 main'' region, \citet{bal14} paired the oppositely facing bow shocks as well as their HH counterparts to the north and south of the strongest clump in the whole region, and designated as MHO~3417, which spans $\sim$10\arcmin\ in total. Unlike the HH features that often extend for many parsecs, molecular outflows are generally observed close to the driving source \citep{bal16}. We found that the blueshifted lobe of DMOC-0040, coupled with a weak redshifted one, is associated with the north portion of MHO~3417. On the other side, although the redshifted lobe of DMOC-0036 coincides with the southern jet, its gas component is blended with another outflow known as MHO~3418 driven by the southwest clump. Observations with higher resolution are needed to determine the spacial distributions and orientations of these lobes.

The ``SW2 SE'' and ``SW2 S'' regions are two sections of a discontinuous filament. A bipolar outflow, DMOC-0039, is revealed within the easternmost clump on the filament. The outflow extend along the east-west direction, and its blueshifted lobe is associated with HH~956 \citep{arm11}. However, such orientation is inconsistence with that of nearby MHOs, which suggest there are multiple outflows from different driven sources. Another active region on the filament is where it breaks and Class~I YSOs emerge. The blueshifted lobe of DMOC-0032 is elongated and extends from the west end of ``SW2 SE'' to the east end of ``SW2 S''. Such structure is likely to be unresolved outflows since it is associated with two groups of H$_2$ knots (MHO 3421/3422) next to two clusters of YSOs. DMOC-0028 is a bipolar outflow associated with a small clumps to the north of the ``SW2 S'' region. The outflow presents clear line wing structures with rather strong emission and large average velocity, but the intensities of dense gas and dust in the clump are relatively weak.

\citet{bal14} found a long chain of H$_2$ shocks through the ``SW1'' region which could be driven by a dim Class~0 protostar in the most opaque portion of the clump. However, no outflow is identified by our procedure. The shocks present large aspect ratio in distribution and clear bow shock features, which suggests that the inclination of the outflow could be large and most of the high velocity components are parallel to the plane of the sky. On the other hand, after carefully checking the spectra, we find several velocity components converging near this region, which prevents the detection of low velocity line wings.

The number density of shock features in the ``core'' region presents a clear enhancement comparing with other subregions. \citet{tou11} pointed out that this subregion has a higher star-formation efficiency. Four molecular outflows lie along the filamentary clouds, which seems less scattered than the distributions of shock features. DMOC-0061 is a pair of lobes associated with a YSO clustered clump near the center of this region. Both lobes spatially coincide with the molecular component at $\sim$5~km~s$^{-1}$, and present similar outflow average velocity. However, due to the complicated distribution of the YSOs, we can not rule out the possibility that the lobes are driven by different sources. A bipolar outflow has been identified in ``core E'' by \citet{gre11} with CO $J=2-1$ observations. \citet{dun12} using their SMA observations confirmed such result, and further concluded that it is ejected by a Class 0 source embedded in the millimeter source MM3 rather than the FU Ori candidate HBC722 nearby. The redshifted lobe, DMOC-0063, is $\sim$1\arcmin\ to the northeast of ``core E'' in our observations, while the blueshifted counterpart is missing. The position of the red lobe matches the orientation of the outflow and the distributions of associated MHOs \citep{dun12, bal14}.

\subsubsection{Pelican's Head}

\begin{figure*}[htbp]
\centering
\includegraphics[width=0.75\linewidth]{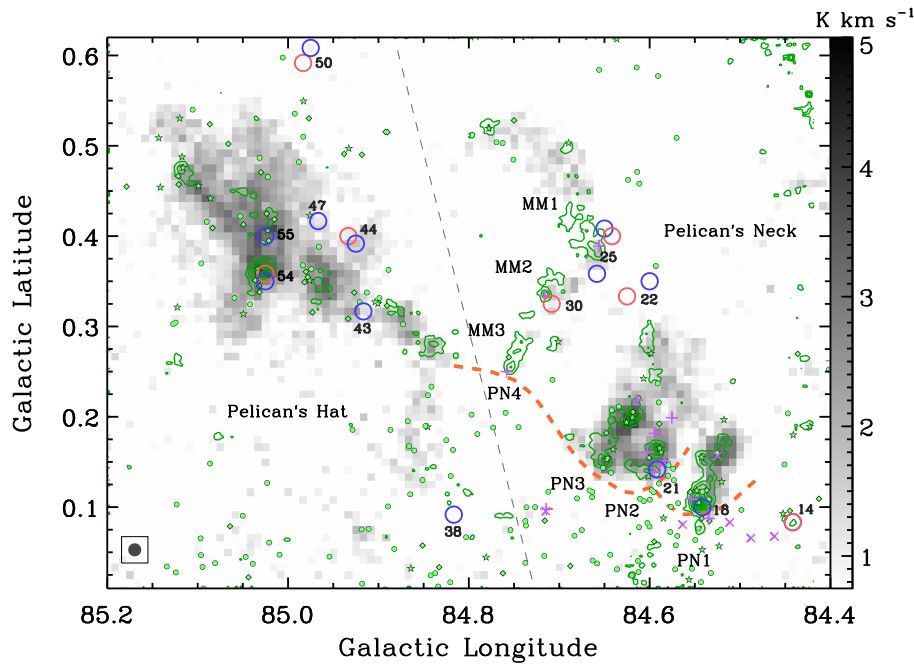}
\caption{Outflows in the Pelican's Head. The background image, contours, and symbols are the same as those in Figure~\ref{fig:regionGM}. The red dashed lines outline the fluorescent edge of H$_2$ emission \citep{bal14}. \label{fig:regionPN}}
\end{figure*}

The Pelican's Head region lies in the northeastern tip of the Pelican Nebula (IC~5070). As shown in Figure~\ref{fig:regionPN}, molecular clouds in Pelican's Head could be subdivided into two sub-regions: Pelican's Hat to the east, and Pelican's Neck to the west. In Pelican's Neck, there are four cometary clouds (designated as PN1-PN4) extend to the north behind the UV-excited edges mentioned by \citet{bal14}. PN4 could be further subdivided into three condensations designated as MM1-MM3. 

The blueshifted lobe DMOC-0018 associates with the strongest clumps of PN1, which is surrounded by MHOs and HH objects. Such outflow could be driven by an Class I YSO, J205041.02+441808.1, embedded in the clump identified by \citet{reb11}. Several shock features emerge around the west two clumps in PN2, but only the blueshifted lobe of DMOC-0021 is detected in the weaker one which is associated with reflection nebula \citep{bal14}. Molecular emission in the stronger clump to the north blends with a weak component at 5.9~km~s$^{-1}$, which prevents us from extracting blueshifted line wing. Shock features emerge in each clump of PN4, while molecular outflows are only found in the northern two. The outflow associated with MM1, DMOC-0025, is a bipolar one, whose blueshifted lobe contains two peaks. Its orientation well matches the position angle of the H$_2$ knots \citep{bal14}. DMOC-0030, coinciding with MM2, shows a prominent redshifted lobe but no localized blueshifted line wing. \citet{bal14} reported a X-shaped outflow (MHO~3400) embedded in the dust clump, which is interpreted as precessing jets of binary.

Molecular emission in Pelican's Hat present an elongated structure, most portion of which is infrared dark in Spitzer 24~$\mu$m image. Several dusty clumps lie in the filament with similar spatial interval of $\sim$~3.5\arcmin. Outflows in Pelican's Hat are relatively weak, with size and mass below the average.
DMOC-0044 is a bipolar outflow without YSO association as driving source. The X-ray observations conducted by \citet{dam17} provide us other possible driving candidates which could be interpreted as weak-line T-Tauri stars missed by YSO searches. In the strongest dust core of this subregion, we detected another bipolar outflow, DMOC-0054, which presents prominent blueshifted lobe. Its redshifted counterpart is weak and may be contaminated by a H~{\scriptsize II} region, $\sim$1\arcmin\ to the northeast of the outflow.

\subsubsection{IC~5146}

\begin{figure*}[htbp]
\centering
\includegraphics[width=0.75\linewidth]{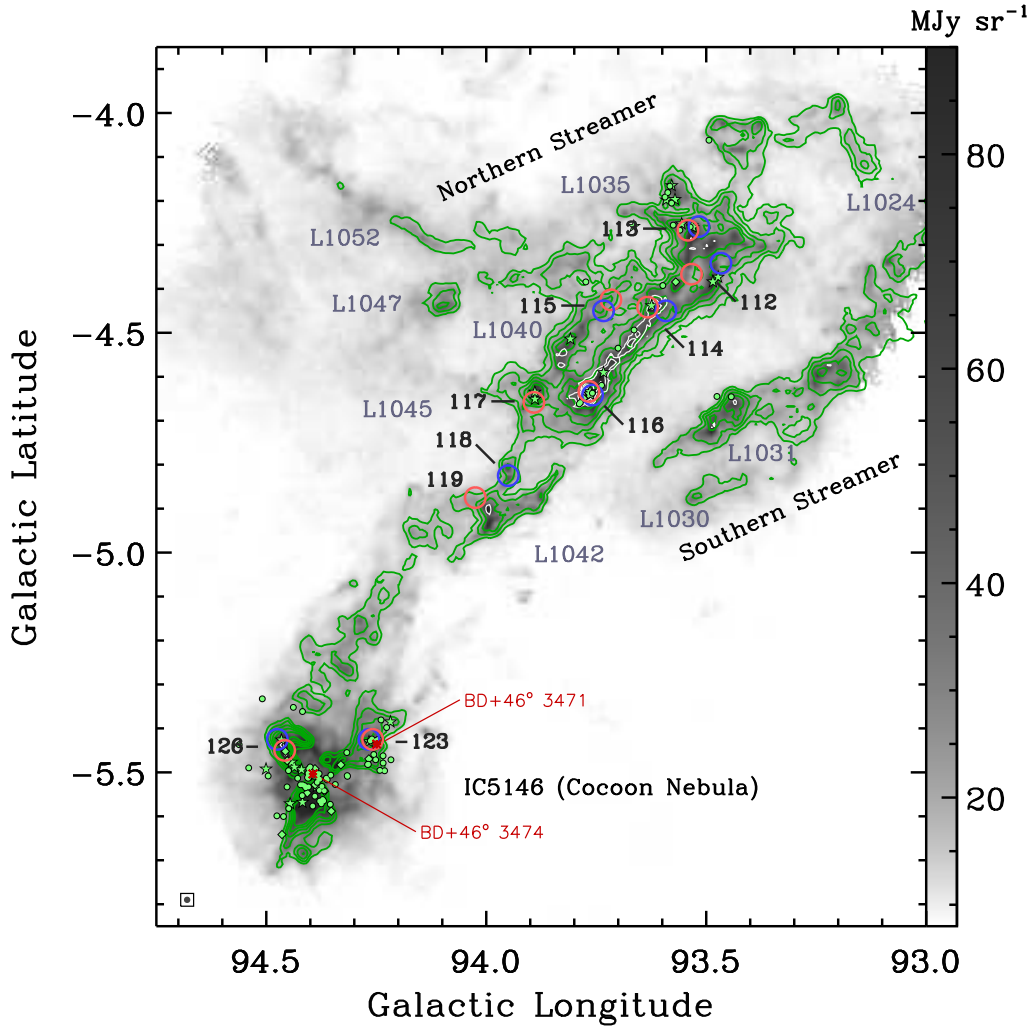}
\caption{Outflows in IC~5146. The gray-scale background is SPIRE 500~$\mu$m image with overlaid contours showing the integrated intensity of $^{13}$CO (green) and C$^{18}$O (white). The spectra are integrated over $-$5 to 15~km~s$^{-1}$. The contours are from 1.6~K~km~s$^{-1}$ ($\sim3\sigma$) and 1.1~K~km~s$^{-1}$ ($\sim2\sigma$) at intervals of $3\sigma$ and $1\sigma$ for $^{13}$CO and C$^{18}$O, respectively. The symbols are the same as those in Figure~\ref{fig:regionGM}. \label{fig:regionIC5146}}
\end{figure*}

IC~5146 consists of clumps and filamentary structures across the Cocoon Nebula and the two Streamers as shown in Figure~\ref{fig:regionIC5146}. The CO emission shows that the Cocoon Nebula is connected to the long filament in the Northern Streamer, while the Southern Streamer which appears to be parallel with the Northern one is attached to the opposite end of it through several sub-filaments. Aside from their spatial connections, the velocities of the three subregions in IC~5146 are consistent, which provides strong evidence for their co-distant property \citep{dob93}. The existence of variable stars, young pre-main-sequence stars, Hα-emission stars, and 200 YSO candidates \citep{her02, hav08} is indicative of active star formation within the region. 

The Cocoon Nebula separates the surrounding molecular cloud into three pieces. DMOC-0126 is associated with a clump at the top of the northeast arc. Both of its lobes are marginally spotted and are assigned as grade B, whereas a C$^{18}$O dense clump is detected just between them. The strong bipolar outflow, DMOC-0123, lying in the northwest piece was first discovered by \citet{lev85}. It spatially coincides with a Herbig Be star, BD$+46\arcdeg~3471$. However, the ZAMS distance of pre-MS star is 355~pc, which is inconsistent with the cloud distance ($\sim$0.9~kpc) we derived and those in the literatures \citep{wal59, eli78, har08}. Thus the outflow candidate is either driven by the pre-MS star with a suspected distance \citep{har08, joh17}, or ejected from other YSO candidates.

\begin{figure}[htbp]
\centering
\includegraphics[width=0.9\linewidth]{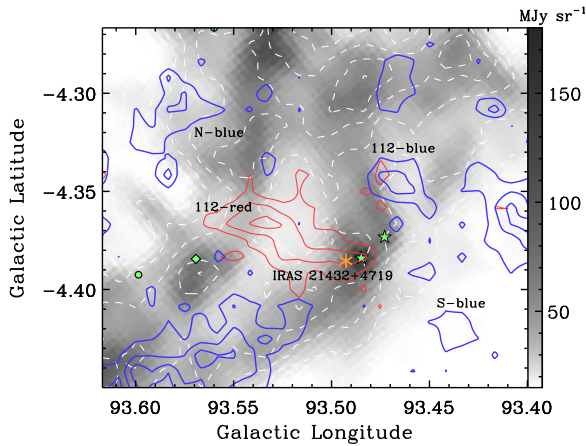}
\caption{CO line wing emissions at DMOC-0112. Blue/red contours are integrated intensity of $^{12}$CO over $-$1.6 to 0.3~km~s$^{-1}$ and 7.6 to 11.9~km~s$^{-1}$. White contours are the integrated intensity of $^{13}$CO over 0.3 to 7.6~km~s$^{-1}$. The symbols are the same as those in Figure~\ref{fig:regionGM}. \label{fig:DMOC-0112}}
\end{figure}

Unlike the Southern Streamer, which shows no sign of any line wing, Northern Streamer possesses 8 outflow candidates in its two parallel sub-filaments and their junctions. DMOC-0116 is a known outflow reported by \citet{mye88}. DMOC-0112, 113, 114, and 117 were firstly identified by \citet{dob93} to be associated with four IRAS sources. Their results are reproduced by our observations in the aspect of position, morphology, and line profile of the lobes. As shown in Figure~\ref{fig:DMOC-0112}, \citet{dob93} identified a massive redshifted lobe, and two marginally resolved blueshifted counterparts designated as S-blue, N-blue near IRAS~21432+4719. Apart from recovering the lobes, a new blueshifted one is detected to the west which is out of the mapping coverage of \citet{dob93}. S-blue seems to be symmetric with the redshifted lobe, but the new blueshifted one is stronger and closer to it. This is indicative of unresolved multiple outflows within the beam, and observations with higher resolution are needed to resolve their origins. As for the N-blue, it is highly possible that it originates from other dense clumps in another sub-filament.

\subsubsection{G084.9$-$0.4}

\begin{figure}[htbp]
\centering
\includegraphics[width=0.8\linewidth]{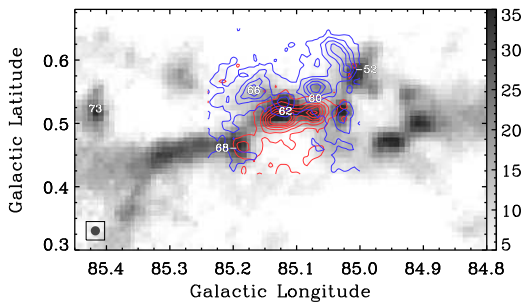}
\caption{CO line wing emissions in the northern arc of G084.9$-$0.4. Background image is the integrated intensity of $^{13}$CO. Blue/red contours are integrated intensity of $^{12}$CO line wing over $-48.5$ to $-44$~km~s$^{-1}$ and $-36$ to $-30$~km~s$^{-1}$.
\label{fig:regionNarc}}
\end{figure}

The G084.9$-$0.4 region possesses 16 outflow candidates which make it one of the most active region in the Perseus arm. The velocity consistency of the rings ($\sim$-40~km~s$^{-1}$) suggests they are at the same distance. Though most molecular gas gathers in the southern ring, only 5 outflows are scattered in it. The most active cloud where outflows crowd is the broken arc to the north as shown in Figure~\ref{fig:regionNarc}. There are 5 candidates arranging in the middle portion of a filamentary cloud. Though it may be ambiguous to confirm outflows in a distant spiral arm, the bipolarity and distinct distributions between blueshifted and redshifted lobes illustrate these line wings originate from outflows rather than other shock features, such as H~{\scriptsize II} region or supernova. The probed lobes, especially those in DMOC-060, 62, and 66, show high mass ($>10~$M$_\odot$) and high velocity($5-10$~km~s$^{-1}$), and their sizes are comparable to the sizes of the associated clumps. There is another outflow, DMOC-0073 lying to the east of the arc, but the velocity of its associated cloud ($\sim$-50~km~s$^{-1}$) indicates that it might be at a further distance.


\subsubsection{Individuals}

$DMOC$-0074\ ($V1057~Cyg$) - this outflow locates to the east of Gulf of Mexico, and is associated with a famous FU ori star, V1057~Cyg. Only the blueshifted lobe has been detected by \citet{lev88}. Its redshifted lobe uncovered in this work is weaker than its counterpart in intensity, but the lobes present similar average velocity and good spatial symmetry.

$DMOC$-0069 - This outflow is the most distant candidate thought its distance estimation may be doubtable. It originates from an isolated molecular cloud. Though both lobes are weak, they have similar properties and good symmetry.

$DMOC$-0077 - Together with DMOC-0069, they are the only two candidates identified in the Outer arm. The outflow is in a filamentary cloud protruded from a giant molecular cloud. Its lobes show similar intensity, but the blueshifted one presents a higher average velocity (4.7~km~s$^{-1}$) than the redshifted one (3.0~km~s$^{-1}$).

\section{Discussion} \label{sec:discuss}

\subsection{Compare Properties with Other Outflow Surveys}

\begin{figure}[htbp]
\centering
\includegraphics[width=0.8\linewidth]{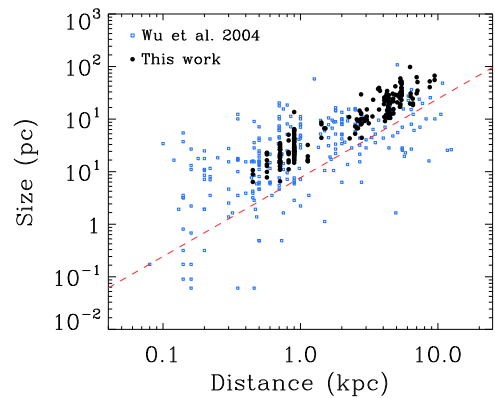}\\
\caption{
The relation between distance and size of the lobes, in which the dashed red line indicates the angular resolution (50$\arcsec$) in this work.
\label{fig:distancesize}}
\end{figure}

\citet{wu04} conducted a large-scale statistical study toward the outflows in the literatures up to then. The properties measured for their samples cover orders of magnitude larger range than ours, and show large scatter. The sizes of the outflows are compared in Figure~\ref{fig:distancesize} as an example. The large scatter could result from their heterogeneous data which include a large variety of outflow tracers, sensitivities, resolutions, and size measurement methods. On the other hand, some scatter would be introduced to our samples if we include more line wings at low velocity that blend with ambient gas as these works. Although the sizes in both catalogs are similar within a distance of $\sim$2~kpc, the distant samples in our work present a slightly larger size than those in \citet{wu04}. Moreover, a similar drop of angular size at larger distance, less severe for our samples, could be spotted, which could be attributed to the limit size of unresolved outflow clusters. The mass, momentum, energy and dynamical time-scale share similar ranges and trends as distance increasing for both catalogs. 

To reduce the discrepancy arisen from different observational and detection methods, we further pay our attention to some unbiased outflow surveys with the same tracer and almost the same resolution and sensitivity. Perseus is one of the regions that molecular outflows have been searched throughly by \citet{arc10}. 60 new candidates are detected using their three-dimensional visualization method. The Taurus region is another well-studied example. \citet{nar12} and \citet{li15} both carried out outflow searching toward the same data set of FCRAO CO survey. Slight biases could be spotted in their methods. \citet{nar12} focused on high-velocity outflows ($>$3~km~s$^{-1}$), and candidates in \citet{li15} are bias to outflows driven by YSOs detectable in Spitzer infrared photometry, which means those associated with Class 0 and other embedded objects are missing. Another two unbiased studies, which are based on the MWISP large scale CO survey as well, are carried out by \citet{li18} and \citet{li19}. They found 198 and 459 outflow candidates toward the Gemini OB1 region and W3/4/5 Complex, respectively.

\begin{figure*}[htbp]
\centering
\includegraphics[width=0.8\linewidth]{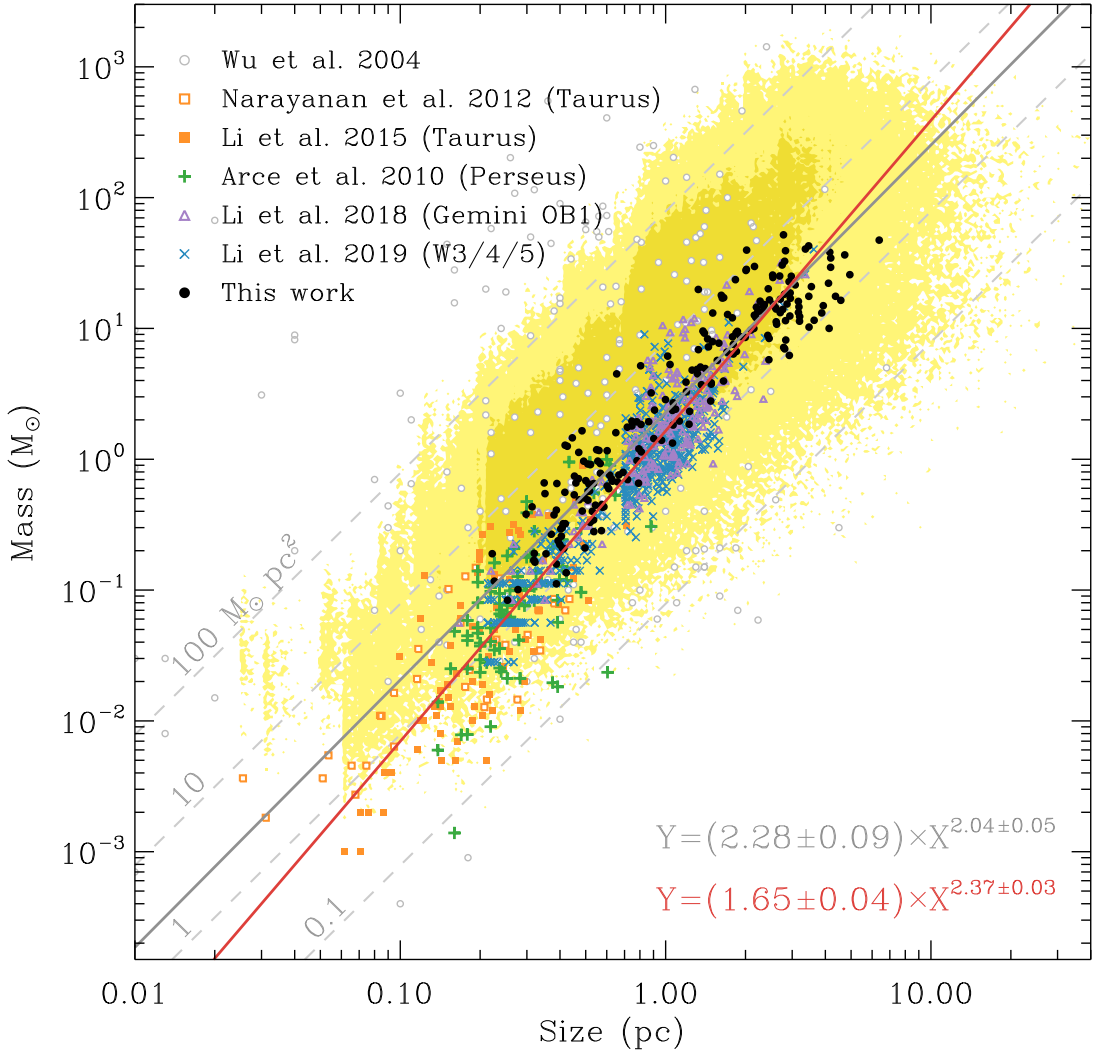}\\
\caption{
Correlation between the masses and sizes of the lobes. Outflow samples in different literatures are shown in different colors and symbols. Yellow filled contours represent the number density of samples after corrections of inclination, opacity, excitation temperature, and low-velocity outflow component using Monte Carlo experiments. The red solid line is the linear fitting result to all samples except those in \citet{wu04}, while the gray solid line is the fitting result only to our work. The dashed gray lines indicate surface density of 0.1, 1, 10, and 100~$M_\odot$~pc$^{-2}$.
\label{fig:massvssize}}
\end{figure*}

We extract and compare two fundamental properties, the lobe size and mass, which are commonly measured in the aforementioned literatures, but the methods used to measure the properties vary among different studies. The lobe sizes estimated at different contour levels are unified to the size at half-maximum contour by simply assuming a Gaussian column density profile, which means a size measured at 30\% contour is multiplied by a factor of 0.76 for comparison. As for the mass, \citet{arc10} adopted a method different from other studies in column density calculations to correct opacity. We manage to recalculate the masses of their samples following the method we used in Section~\ref{sec:properties}. Other parameters such as excitation temperature, CO abundance, and mean molecular weight are also unified. Eventually, the correlation between the lobe sizes and masses is shown in Figure~\ref{fig:massvssize}.

A tight power-law correlation appears in the plot for the unbiased studies, while such correlation is not significant in the samples from \citet{wu04} due to large scatter. It is worth noting that properties of the points in Figure~\ref{fig:massvssize} are not corrected for the facts of inclination, opacity, excitation temperature, low-velocity outflow component, and high-velocity outflow emission below the sensitivity. We managed to correct the first four factors with Monte Carlo experiments. The correction factor and method are mentioned in Section~\label{sec:properties}. After adding factors for 500 experiments, we use filled yellow contours to illustrate the number density of the expanded results in Figure~\ref{fig:massvssize}. Larger intercept and extra scatter are introduced to the correlation which is, however, still remain. The correction for outflow emission hidden below sensitivity is difficult to estimate, and may affect the slope of the correlation, since masses in distant outflows are more likely to be underestimated due to beam dilution. More observations combining with theoretical studies are needed to measure all correction factors comprehensively. The correlation between size and mass is usually interpreted as self-similar fractal structures found in interstellar medium (ISM) on different spatial scale spanning from giant molecular clouds (GMCs) to cores \citep{elm96, rom10, kau10, lom10a}. Such self-similar also emerges in filamentary molecular clouds, as a rich structures of sub-filaments are commonly resolved in MWISP filament surveys \citep{xio17, xio19} and higher-resolution studies \citep{hac18}. \citet{man83} pointed that the fractal structure is naturally expected in turbulent ISM. As small scale and high-velocity sub-structures of molecular clouds, low-J CO outflows could reflect the structures of molecular gas around the driven sources. Our correlation indicates that the self-similar structures of ISM may persist in outflows from the scale of outflow clusters to individual cavities, though jets may replace certain roles of turbulence in them.

When discussing fractal in ISM, the fractal dimension is usually used to describe how turbulent gas of small scale fills the upper level space, and forms self-similar hierarchical structures. A similarly defined fractal dimension \citep{rom10}, $D_M = 2.37\pm0.03$, could be derived by applying an ordinary least-squares bisector linear fitting between the logarithms of sizes and masses shown in Figure~\ref{fig:massvssize}. Though such mass-based fractal dimension is a projected result of ISM in three-dimensional space, a simulation study carried by \citet{san05} found that the value of three-dimensional fractal dimension is quite close to $D_M$. A fractal dimension over 2 indicates outflows or their clusters are complicated spongy structures filling the three-dimensional spaces rather than simple sheets. Our fractal dimension is quite similar to the exponent derived from the molecular cloud surveys \citep{rom10, col19}, as well as several GMCs \citep{elm96}, which is over pc scale dominated by turbulence, but smaller than the case of M17 where a bright H~{\scriptsize II} region provides extra ambient pressure to the molecular clouds \citep{stu90, elm96}. Another measurement using near-infrared extinction map presented by \citep{lom10a} reported an smaller exponent of $\sim$2.0 on pc scale and a drop to 1.6 on sub-pc scale, which is agree with the one obtained by \citet{kau10}. The higher fractal dimension in outflows might coincide with the simulation results reported by \citet{whi13} that outflows raise the clumpiness in the clumps, and may suggest that the outflows contain more turbulent motions than expected.





\subsection{The Impact of Molecular Outflows}

The large-scale searching for outflows could help us to assess the transferring process of material, momentum, and energy from outflows to their parent clouds. Table~\ref{tab:feedback} shows the total mass, momentum, and energy the outflows carry, along with the gas mass, turbulent momentum and energy the molecular clouds possess in seven regions. Though the total mass of the regions differ within an order of magnitude, the proportion of their outflowing gas is $\sim$1.4\%. Similar situations are found in the outflows contributing to the turbulent momentum ($\sim$1.3\%) and energy($\sim$0.8\%). Even if we consider the correction for optical depth and inclination, the ratio $E_{\rm flow} / E_{\rm turb}$ is less than $40\%$. The low ratio indicates that outflows are insufficient to energize cloud-scale turbulence solely. The incapability of outflow has been found by several literatures \citep{arc10, li15, li18}. When further noting that outflow is not detected in numerous quiescent clouds, it is evident that other turbulence energy injection mechanisms such as H~{\scriptsize II} region, feedback from supernova, and gravitational instability \citep{kru16} are working. Simulations and principal component analysis of NGC~1333 region conducted by \citet{bru09} also support that cloud-scale turbulence may be driven by large scale injection of energy outside the cloud.

\begin{deluxetable*}{cclcDDDlcDDD}
\addtolength{\tabcolsep}{-1pt}
\tablecaption{Feedback of outflows \label{tab:feedback}}
\tablewidth{0pt}
\tablehead{
\colhead{} & \colhead{} &
\colhead{} &
\multicolumn{7}{c}{Outflow} &
\colhead{} &
\multicolumn{7}{c}{Cloud}
\\
\cline{4-10} \cline{12-18}
\colhead{Name} & \colhead{Distance} &
\colhead{} &
\colhead{n} & \multicolumn2c{$M_{\rm flow}$} & \multicolumn2c{$P_{\rm flow}$} & \multicolumn2c{$E_{\rm flow}$} &
\colhead{} &
\colhead{Area} & \multicolumn2c{$M_{\rm cloud}$} & \multicolumn2c{$P_{\rm turb}$}& \multicolumn2c{$E_{\rm turb}$}
\\
\colhead{} & \colhead{(kpc)} &
\colhead{} &
\colhead{} & \multicolumn2c{(M$_\odot$)} & \multicolumn2c{(M$_\odot$ km s$^{-1}$)} & \multicolumn2c{($10^{44}$ erg)} &
\colhead{} &
\colhead{(pc$^2$)} & \multicolumn2c{($10^3$ M$_\odot$)} & \multicolumn2c{($10^3$ M$_\odot$ km s$^{-1}$)} & \multicolumn2c{($10^{47}$ erg)}
}
\decimals
\startdata
L931\tablenotemark{a} & 0.71 & &  6 &  5.2 &  \quad\ 19.2 &  0.8 &   & 4.3 & \quad 1.0 & \qquad\ 3.1 & \quad 2.1 \\
Gulf of Mexico  & 0.60 & & 28 & 34.4 & 198.5 & 12.6 &   & 5.8 &  5.7 & 21.2 & 10.8 \\
Pelican's Neck  & 0.60 & & 13 & 50.9 & 195.5 &  8.8 &   & 3.5 &  3.2 & 11.8 &  7.3 \\
Pelican's Hat   & 0.60 & & 10 & 16.9 &  60.7 &  2.6 &   & 4.6 &  1.6 &  5.2 &  2.7 \\
L941            & 0.45 & &  6 & 26.1 &  89.7 &  3.4 &   & 2.5 &  0.7 &  4.7 &  5.2 \\
IC5146 Streamer & 0.45 & & 14 &  6.5 &  29.4 &  1.5 &   & 2.5 &  1.5 &  4.0 &  3.2 \\
IC5146 Cocoon   & 0.45 & &  4 &  6.1 &  13.9 &  0.4 &   & 0.4 &  0.3 &  0.7 &  0.5 \\
\enddata
\tablecomments{Comparison of properties between outflows and molecular clouds. The first two columns list the name and distance of the regions. The ``Outflow'' section gives the number, total mass, momentum, and energy of lobes. The ``Cloud'' section includes the area within the half-maximum contour line of $^{13}$CO emission, mass estimated using $^{13}$CO under the LTE assumption, momentum and energy contained in turbulent motions.}.
\tablenotetext{a}{L931 is not fully covered in our observations, thus the parameters are lower limits.}
\end{deluxetable*}

Though outflows could not provide turbulent energy on cloud-scales, they may still affect their immediate environment. When we focus our study to the candidates in the Local arm, we could simple estimate the spatial scale, in which turbulent energy in ambient gas is equivalent to the energy of a lobe, to be $\sim$0.12~pc, clearly smaller than the clump size that \citet{zha14} measured in the North American and Pelican Nebulae. Such scale indicates the impact of outflows is restricted within the size of cores \citep{ber07}. However, the estimation is rough since we need to further consider the timescale of energy injection and dissipation, since the turbulent energy dissipates rapidly in molecular clouds \citep{mac04}. \citet{mac99} related the timescale of turbulence dissipation to the free fall time in their numerical computations, and gave $t_{\rm diss} / t_{\rm ff} \simeq 3.9~\kappa / M_v$, where $\kappa$ is the ratio between turbulence driving length and Jeans length, and $M_v$ is the Mach number of turbulence. The driving length of a continuous outflow is approximate to the lobe length as shown in numerical simulations \citep{nak07, cun09}. By using a gas temperature of $\sim$10K and mean density of $1.1\times 10^4$~cm$^{-3}$, we could derive a sound speed of 0.3~km~s$^{-1}$, a Jeans length of 0.3~pc, and a free fall time of 0.36~Myr. We could then deduce a dissipation timescale around 0.31~Myr, which is larger than the mean dynamical timescale of outflow by a factor of 2.5. Thus after considering the timescale, the spatial scale, in which the dissipation rate of turbulent energy is equivalent to the energy injection rate of a lobe, turns into $\sim$0.19~pc. The result confirms the prediction of \citet{bru09} that outflow is important on providing turbulent energy on core- or clump-scales on short time scales.


\subsection{Missing Outflow Candidates}

Within our mapping region, two outflows reported in the literatures are not detected. An instance, associated with molecular core B361 \citep{bei86}, was found by \citet{wu92} in their CO~(2$-$1) survey. No sign of line wing could be found in our J=1$-$0 transition. The other one near IRAS 21428+4732 was revealed by \citet{dob01} in their CO~(1$-$0) observations with 45-m telescope at Nobeyama Radio Observatory. Due to beam dilution, only the blueshifted lobe is marginally detected in our observations. Thus as a mono-polar grade B sample, it is not included in our catalog. The failure in detecting these two sources reminds us the possible factor that may affect the result of outflow searching.

The tracer, observational sensitivity and resolution are the dominant factors that limit the amount of protostellar outflows detected in molecular cloud, but it is difficult to estimate how many outflows may be missing. We looked up the record of YSOs and detected molecular outflows in two nearby star forming regions, the Perseus and Taurus region, from the literatures. The amount of detected molecular outflows is a factor of $\sim$7 less than the number of corresponding YSOs \citep{eva09, arc10, cur10, reb10, li15}. For NGC~1333, a subregion in Perseus, \citet{bal16} pointed out that the number of known flows is less than that of YSOs by a factor of at least 4. If we adopt the factor of 4 to the North America and Pelican Nebulae where YSOs are found in detail by \citet{reb11}, there should be $>$60 outflows in the Gulf of Mexico or the Pelican region. Such estimated amount is 4 times more than the outflows we actually detected. An obvious reason is that most outflows are not resolved, especially in the outflow clusters, since our spatial resolution ($\sim$0.15~pc) is 2-4 times as large as those in Perseus and Taurus. Another important reason is the different compositions of YSOs, since the outflow detection rates vary in different classes of YSOs \citep{li15}.


Other than the observational factors, outflow searching methods also determine the results, and may accidentally miss some particular samples. Most outflow studies based on low-J CO (\citealt{arc10, li15}; \citealt{li18} and this work) use the combination of $^{12}$CO and $^{13}$CO to probe line wing and cloud structure, respectively, which means that similar features emerged on $^{13}$CO or C$^{18}$O lines are usually missing. The detection rate might be reduced where $^{12}$CO line width are heavily broadened or multiple velocity components are blended. On the other hand, unlike higher J transitions of CO which are better at tracing high velocity component of outflow, J=1$-$0 transition requires the detection of $^{13}$CO to estimate the velocity dispersion in ambient gas. Consequently, outflow can not be detected if there is no associated $^{13}$CO emission. Moreover, to improve the reliability of detection, we discard the mono-polar grade B and all grade C samples, since they could be highly contaminated by other kinematic structures such as expanding H~{\scriptsize II} regions, supernova remnants, fragmentation in clumps, velocity gradient on filaments, and even emission from ambient gas. And as a result, we can not rule out the possibility that weak outflows are within them, and inevitably miss them.



\section{Summary} \label{sec:summary}

We have conducted a survey of molecular outflows for a complex of dark clouds in the Cygnus region using $^{12}$CO, $^{13}$CO, and C$^{18}$O molecular line maps from MWISP survey. Our 46.75~deg$^2$ maps cover a large number of cold dark clouds, and provide abundant data to study outflow properties and feedback during the formation of stars. We developed a searching method that combines the machine learning algorithm, SVM, with traditional visual inspection to efficiently recognize outflow line wing features in position-position-velocity space. Our main results are summarized as follows.

\begin{enumerate}
\item A total of 130 outflow candidates are identified within the mapping area, and 118 of them are new detections. Outflows usually emerge in dense molecular clouds, and form clusters or linear structures along molecular filaments according to their environments. 97, 31, and 2 candidates locate in the Local, Perseus, and Outer arms, respectively.

\item 59.2\%, 22.3\%, 18.5\% of the outflows in our catalog are bipolar, mono-blue, and mono-red polar, respectively. The mean size of our outflows is 0.54~pc, and the mean mass is 0.72~M$_\odot$ in the Local arm. Some outflows show shock indicators such as HH objects, MHOs, and water masers. Most outflows may be driven by YSOs, while there are still 36 candidates not associated with any detectable YSO.

\item Molecular clouds and clusters of outflows are revealed in several star forming regions, which reflect diverse level of star forming activities. Outflows shown by CO also present different distribution from that traced by shock features.

\item The outflow properties we measured agree with those in the literatures. A tight power-law correlation between the lobe sizes and masses appears when comparing samples of different works, which suggests self-similar structures emerge in outflows. A higher fractal dimension of $2.37\pm0.03$ than that in clumps indicates that turbulence brought by outflows may raise the clumpiness in molecular clumps.

\item By comparing energy injected by outflows and turbulent energy on different spatial scales, we note that outflow is insufficient to energize cloud-scale turbulence, but may be important on support turbulence on core- or clump-scales on short time scales.
\end{enumerate}


\acknowledgements{This work is based on MWISP data acquired with the Delingha 13.7 m telescope of the Purple Mountain Observatory. The authors appreciate all the staff members of the observatory for their help with the acquisition and reduction of the data. Our gratitude also goes to the MWISP team for their support of operating the project. This work is supported by National Key Research \& Development Program of China grant no. 2017YFA0402700, Key Research Program of Frontier Sciences, CAS, grant no. QYZDJ-SSW-SLH047, the National Natural Science Foundation of China through grants NSF~11803091, 11873093, 11773077, 11873019, 11673066, and by the Key Laboratory for Radio Astronomy, CAS.}

\appendix
\restartappendixnumbering
\section{Accuracy and F1 Score}\label{appendix}

When applying a classification algorithm to a dataset pre-labeled with positive and negative tags, four situations would appear in the predictions:
\begin{itemize}
\item True Positive (TP): positive sample predicted as positive
\item False Negative (FN): positive sample predicted as negative
\item False Positive (FP): negative sample predicted as positive
\item True Negative (TN): negative sample predicted as negative
\end{itemize}
A straightforward metric to evaluate the performance of the classifier is the accuracy:
\begin{displaymath}
\rm Accuracy = (TP + TN) / SampleSize
\end{displaymath}
However, accuracy becomes insensitive in imbalanced classes in which positive samples is far less than the negative ones, especially when a study is focusing on the less side. Thus two more metrics are introduced:
\begin{displaymath}
\rm Precision = TP / (TP + FP);
\end{displaymath}
\begin{displaymath}
\rm Recall = TP / (TP + FN);
\end{displaymath}

\begin{figure*}[htbp]
\centering
\includegraphics[width=0.6\linewidth]{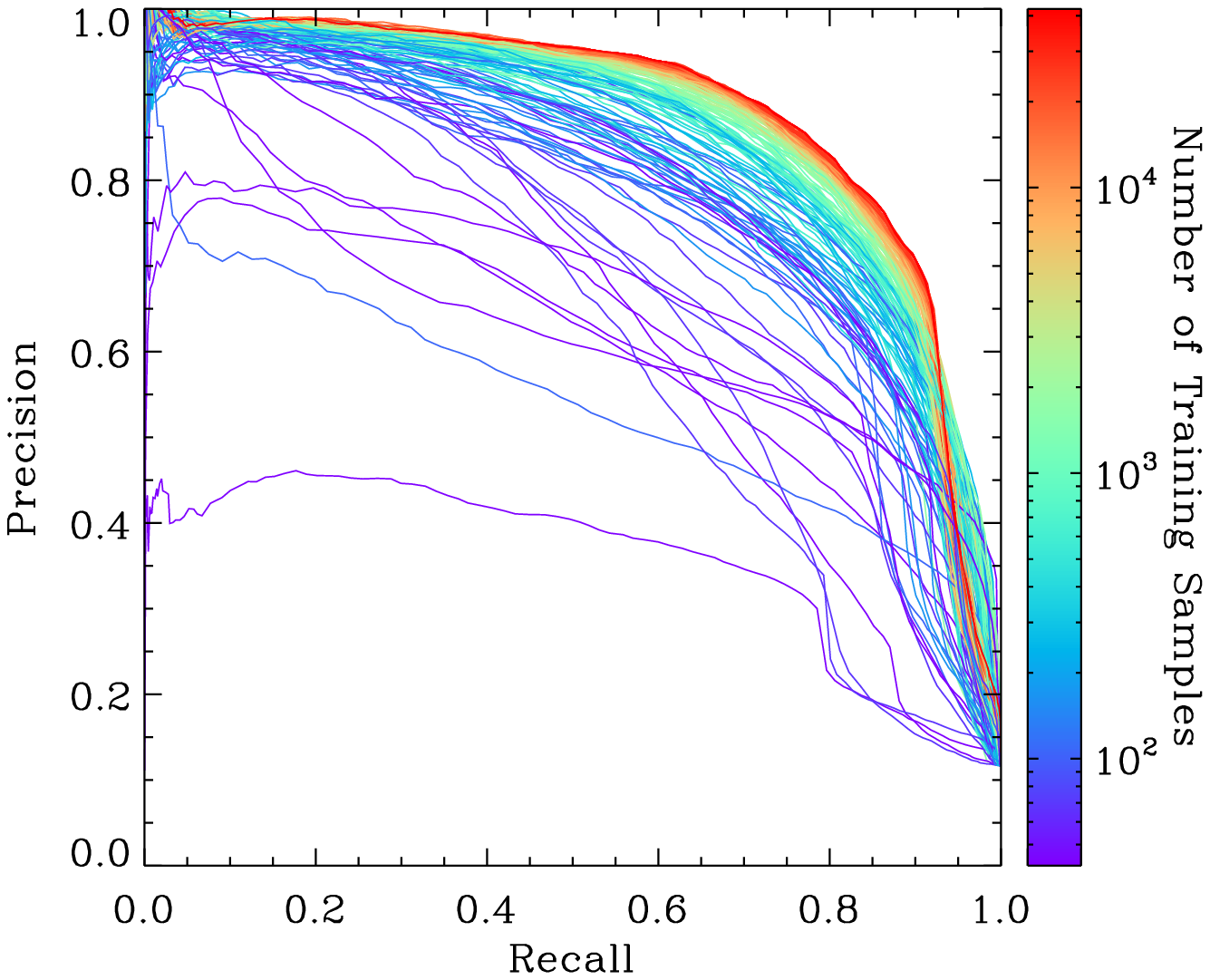}\\
\caption{Trade-off between precision and Recall. Each curve represents a model trained from different sample size. \label{fig:prcurves}}
\end{figure*}

In our outflow searching study, precision indicates the proportion of how many actual LHW pixels (labeled with positive) are among all selected candidates (predicted as positive), while recall indicates the proportion of how many actual LHW pixels are retrieved from the dataset. The relationship between precision and recall for our models is shown in Figure~\ref{fig:prcurves}. The area under the curve represents the overall capability of a model. Apparently, precise model tends to give low recall rate, and precision can not be guaranteed by a model that recalls all samples. To balance the trade-off between precision and recall, F1 score could be defined as:
\begin{displaymath}
\rm F1~score = 2 \times {Precision \times Recall \over Precision + Recall}
\end{displaymath}
which makes it a more useful metric than accuracy to assess classification models for imbalanced problems.

\software{GILDAS/CLASS \citep{pet05, gil13},
SVM$^{light}$ \citep{joa99},
IDL Astronomy Library \citep{lan93astrolib}}

\bibliographystyle{aasjournal.bst}
\bibliography{reference}{}

\end{document}